\documentclass[12pt]{article}
\usepackage{amssymb}
\usepackage{amsfonts}
\usepackage{amsmath}
\usepackage{amsmath,amsthm}
\usepackage{graphicx,color}
\begin{document}
\begin{center} {{\bf {Reissner-Nordstr\"{o}m black holes statistical ensembles\\ and\\ first order thermodynamic phase transition
}}\\
 \vskip 0.25 cm
  {{  Hossein Ghaffarnejad\footnote{E-mail address: hghafarnejad@semnan.ac.ir }
   and Mohammad Farsam \footnote{E-mail address: mhdfarsam@semnan.ac.ir}}}}\vskip 0.2 cm \textit{ Faculty of Physics, Semnan
University, Zip Code: 35131-19111, Semnan, Iran}
\end{center}
\begin{abstract}
We apply Debbasch proposal to obtain  mean metric of coarse
graining (statistical ensemble) of quantum perturbed
Reissner-Nordst\"{o}m black hole $(RNBH).$ Then we seek its
thermodynamic phase transition behavior. Our calculations predict
first order phase transition which it can take Bose Einstein's
condensation behavior.
\end{abstract}
\section{Introduction}
Every observation in any arbitrary system is necessarily finite
which deals with a finite number of measured quantities with a
finite precision. A given system is therefore generally
susceptible of different, equally valid descriptions and building
the bridges between those different descriptions is the task of
statistical physics (see introduction in ref. [1] for more
discussion). Nonlinearity property of Einstein`s metric equations
cause to be nontrivial their averaging. Various possible ways of
averaging the geometry of space time have already been proposed by
[3-9], but none of them seems fully satisfactory (see section 7 in
ref. [1] for full discussion). Debbasch is used an alternative way
to averaging the Einstein`s metric equation in [1]. To do so he
chose a general framework where the mean metric is still obeys the
equations of general theory of relativity. In his approach
averaging and/or coarse graining a gravitational field changes the
matter content of space time called as `apparent matter` which in
cosmological context is related to the dark energy (see [10-13]).
So general relativity mean field theory can propose a physical
meaning for un-known cosmological dark energy/matter via the
`apparent matter`. In the Debbasch approach, statistical ensemble
of metric is ensembles of histories and not ensembles of states.
This is different basically with ordinary statistical mechanics of
classical and/or quantum particles. From the latter point of view,
it has been known for a long time that black holes in
asymptotically flat space-times do not admit stable equilibrium
states in the canonical ensemble (see introduction in ref. [14]).
But from the former point of view the Debbasch gives in ref. [1],
general proposal to obtain a mean field theory for the general
theory of relativity. In his model members of the ensembles will
be labeled by the symbol $\omega\in\Omega$ where $\Omega$ is an
arbitrary probability space [15]. To each $\omega,$ there are
corresponding metric tensor $g(\omega),$ compatible connections
$\Gamma(\omega)$ and the Einstein metric equation (see [1] and
section 2 in ref. [11]). All members of the ensemble correspond to
the same macroscopic history of the space time manifold, in
particular to a  given same mean metric
$\bar{g}_{\mu\nu}(x)=<g_{\mu\nu}(x,\omega)>$ and corresponding
mean connection
$\bar{\Gamma}^{\mu}_{\nu\eta}=<\Gamma^{\mu}_{\nu\eta}(g,\partial
g, \omega)>$. As application of his model Debbasch and co-workers
considered statistical ensemble of Schwarzschild black holes as
non-vacuum solutions of mean Einstein metric equation by using
Kerr-Schild coordinates $R=r-\omega.$ They calculated
non-vanishing temperature of mean metric where single
Schwarzschild black hole is well known that has non-vanishing
temperature as a vacuum solution of the Einstein equation.  They
discussed their results with special emphasis on their connections
with the context of astrophysical observations [13]. Extreme RNBH
with $m=1$ has vanishing temperature (see next section) and
regular Kerr-Schild coordinates $R=r-\omega$ are not applicable to
obtain mean metric similar to the Schwarzschild one because the
coarse graining space-time turns out not to be a black hole [10].
Hence Chevalier and Debbasch used analytic continuation of the
Kerr-Schild coordinates as $R=r-i\omega$ to obtain mean metric of
extreme
 classical black hole in ref. [12]. According to the Debbasch approach  we are
free to choose types of coarse graining and/or ensemble space to
obtain mean metric of the space times ensemble under
consideration. We should point that topology of ensemble space
times must be similar to topology of their mean metric (see ref.
[10]) which restrict us to choose an analytic continuation of
Kerr-Schild coordinates for extreme RNBH. In short, with Debbasch
proposal the averaging process dose not change topology between
ensemble of the curved space-times and the corresponding mean
space time. Precisely that the averaging process modifies the
horizon radius and changes the energy-momentum tensor of
space-time but not total energy or mass of the black holes
ensemble. Really the averaging process just redistributes without
any change in the total mass which means that the total energy of
the black holes dose not
changed by the coarse-graining proposal.\\
Similar to study  thermodynamic behavior of single RNBH [2] we
seek thermodynamic aspect of mean metric of
 non-extreme RNBHs ensemble in this work,
by applying the Debbasch approach to evaluate the mean and/or
coarse
graining metric. Organization of the paper is as follows.\\
 In section 2,
we calculate mean metric of ensemble of RNBHs. In section 3 we
obtain locations of mean metric horizons. In section 4 we
calculate, interior and exterior horizons entropy, temperature,
heat capacity, Gibbs free energy and pressure of RNBHs mean
metric. In section 5 we calculate interior and exterior horizons
luminosity and corresponding mass loss equation of quantum
perturbed RN mean metric. Section 6 denotes to concluding remark
and discussion.
\section{RNBHs ensemble and mean metric}

Exterior metric tensor of a single charged, non-rotating,
spherically symmetric body is given by
\begin{equation}
ds^2=\left(1-\frac{2M}{r}+\frac{{e}^2}{r^2}\right)dt^2-\frac{dr^2}{\left(1-\frac{2M}{r}
+\frac{{e}^2}{r^2}\right)}-r^2(d\theta^2+\sin^2\theta
d\varphi^2).\end{equation} This is metric solution of
Einstein-Maxwell equation and is called as RNBH in which $M$ and
$e$ are corresponding ADM mass and electric charge defined in
units where $c=G=1.$ Equating
$g^{\mu\nu}\partial_{\mu}r\partial_{\nu}r=0$ for arbitrary
spherically symmetric hypersurface $r=constant$ one can obtain
apparent (exterior) horizon radius as $r_+=M+\sqrt{M^2-e^2}$ and
Cauchy (interior) horizon radius as $r_-=M-\sqrt{M^2-e^2}$ which
appear only for $0\leq(e/M)^2\leq1.$ One can obtain mass
independent relation between $r_{+}$ and $r_-$ as
$r_-=\frac{e^2}{r_+}.$ With particular choice $e=M$ (called as
extreme and/or Lukewarm RNBH) these horizons coincide as
$r_-=r_+=M.$ Clearly the RNBH metric solution (1) leads to
Schwarzschild one by setting $e=0$ for which we will have $r_+=2M$
and $r_-=0$. Temperature of a single RNBH can be obtained for
interior and exterior horizons as $T_{\pm}=\frac{1}{8\pi
r_{\pm}}\big(\frac{\partial r_{\pm}}{\partial M}\big)^{-1}_e=
\frac{\pm\sqrt{M^2-e^2}}{8\pi(M\pm\sqrt{M^2-e^2})^2}$ [1]  which
reduce to a zero value for extreme (Lukewarm) RNBH because of
$M=e.$ They show positive (negative) temperature for exterior
(interior) horizons. Negative temperatures of systems have
physical meaning and are happened under particular conditions.
More authors are studied conditions where the physical systems
take to have negative temperatures. See [16] for temperatures of
interior and exterior horizons of Kerr-Newman black hole. One can
see [17,18,19] for negative temperature of non-gravitational
systems. In the nature, materials are obtained which have
interesting properties like negative refraction index,
reversibility of the Doppler`s effect, and so the phase and group
velocity (velocity of energy propagation) have opposite singes. In
these systems temperature will be have negative values (see [17]
and references therein). Such systems are called as dual system
(left-handed) of direct counterpart (right-handed conventional
materials). Absolute temperature is usually bounded to be positive
but its violation is shown in ref. [18] by Braun et al. They
showed under special conditions, however negative temperatures
where high energy states are more occupied than low energy states.
Such states have been demonstrated in localized systems with
finite, discrete spectra. They used the Bose-Hubbard Hamiltonian
and obtained attractively interacting ensemble of ultra-cold
bosons at negative temperature which are stable against collapse
for arbitrary atom number. Furman et al are studied in ref. [19]
behavior of quantum discord of dipole-dipole interacting spins in
an external magnetic field in the whole temperature range
$-\infty<T<\infty.$ They obtained that negative temperatures,
which are introduced to describe inversions in the population in a
finite level system, provide more favorable conditions for
emergence of quantum correlations including entanglement. At
negative temperature the correlations become more intense and
discord exists between remove spins being in separated states.
According to the documentation, and looking to diagrams of the
present work one can be convinced that a quantum perturbed mean
metric of coarse graining RNBHs will be exhibit finally with a
first order phase transition and Bose-Einstein condensation state
microscopically.
 According to the
Debbasch approach [1] ensemble of the non-extreme RNBHs are
collections of coarse graining RNBHs indexed by a 3 dimensional
real parameter $\vec{\omega}\in\vec{\Omega}$ where $\vec{\Omega}$
is the three ball of radius $\vec{a}$ as follows.
\begin{equation}\vec{\Omega}=\{\vec{\omega}\in \mathbb{R}^3; \omega^2\leq a^2\}.\end{equation}
The metric solution (1) is convenient
 to be rewritten with Kerr-Schild
coordinates $(\tau,r,\theta,\varphi)$ by transforming
\begin{equation} dt=d\tau+\frac{h(r)dr}{1-h(r)}\end{equation} as follows (see [11-13]).
\begin{equation}
ds^2= d\tau^2-d\vec{r}\cdot
d\vec{r}-h(r)\bigg(d\tau-\frac{\vec{r}\cdot d\vec{r}}{r}\bigg)^2
\end{equation}
where
\begin{equation}
h(r)=\frac{2M}{r}-\frac{e^2}{r^2}
\end{equation}
and $r=|\vec{r}|$ is the Euclidean norm of the vector $\vec{r}$.
It should be pointed that all metric solutions of the Einstein`s
field equation  will be have simple form by using Kerr-Schild
coordinates. They are decomposed into the well known flat
Minkowski background metric $\eta_{\mu\nu}$ and null vector fields
$K_{\mu}$ as $g_{\mu\nu}=\eta_{\mu\nu}-2h(x^{\mu})K_{\mu}K_{\nu}$
where
$K_{\mu}K^{\mu}=0=g_{\mu\nu}K^{\mu}K^{\nu}=\eta_{\mu\nu}K^{\mu}K^{\nu}$
and $h(x^{\mu})$ is a scalar function (see [20] and references
therein). Now, we must be choose
 a probability measure. Hence we follow the assumption presented in ref.[12] and choose
uniform probability measure $d\textsf{p}_{\omega}$ in which
$\textsf{p}$ is probability density of this measure with respect
to Lebesgue measure $d^3\omega$  as
$\textsf{p}(\omega)=\frac{1}{V_a}$ with $V_a=\frac{4}{3}\pi a^3.$
Applying the Kerr-Schild radial coordinate\footnote{ In case of
extreme RNBH where $M=e$ we must be use analytic continuation of
the Kerr-Schild coordinates as $R=r-i\omega$ (see discussion given
in the introduction).}
$\vec{R}(\vec{r},\vec{\omega})=\vec{r}-\vec{\omega}$ we
 extend single RNBH metric (4) to obtain metric of coarse graining and/or statistical ensemble of
 RNBHs as follows.
\begin{equation}
ds^2= d\tau^2-d\vec{r}\cdot
d\vec{r}-h(R)\bigg(d\tau-\frac{\vec{R}.d\vec{r}}{R}\bigg)^2
\end{equation}
where  $h(R)=\frac{2M}{R}-\frac{e^2}{R^2} $
 and $R=\sqrt{\vec{R}\cdot\vec{R}}.$ Using perturbation series
 expansion method and averaging the metric (6) against
 $\vec{\omega}$ we obtain mean metric of the equation (6) such that  (see [21] for details of calculations)
\begin{equation}
\left<ds^2\right>_{\omega}=b_1(r)d\tau^2+b_2(r)d\vec{r}\cdot
d\vec{r}+b_3(r)dr^2+b_4 (r)dr d\tau
\end{equation}
where $|e|<M,$ $d\vec{r}\cdot
d\vec{r}=dr^2+r^2(d\theta^2+\sin^2\theta d\varphi^2),$
\begin{equation}
b_1(r)=1-\frac{2M}{r}+\frac{e^2}{r^2}\left(1+\frac{a^2}{5r^2}\right),\end{equation}
\begin{equation}
b_2(r)=-1-\frac{2a^2M}{5r^3}+\frac{a^2e^2}{5r^4},
\end{equation}
\begin{equation}
b_3(r)=-\frac{2M}{r}\left(1-\frac{3a^2}{5r^2}\right)+\frac{e^2}{r^2}\left(1-\frac{2a^2}{5r^2}\right),
\end{equation}
 and \begin{equation}
b_4(r)=\frac{4M}{r}\left(1-\frac{a^2}{5r^2}\right)-\frac{2e^2}{r^2}.
\end{equation}
 It is simple to show that the
mean metric (7) reduces to a single RNBH metric (4) by setting
$a=0$. We can rewrite the mean metric (7) in the static frame by
defining the Schwarzschild coordinates. To do so, we first choose
a suitable local frame with coordinates $(t,\rho,\theta,\varphi)$
as
 \begin{equation}\rho(r)=r\sqrt{-b_2(r)}\end{equation} and \begin{equation}d\tau=dt-\alpha(\rho)d\rho\end{equation}
where
 \begin{equation}\alpha(\rho)=\frac{ b_4(r)}{2b_1(r)}\bigg(\frac{\partial\rho}{\partial r}\bigg)^{-1}.\end{equation}
In the latter case the mean metric (7) reads
\begin{equation}
\left<ds^2\right>=F(\rho)d\tau^2-f(\rho)d\rho^2-\rho^2(d\theta^2+\sin^2\theta
d\varphi^2)
\end{equation}
where we defined
\begin{equation}
F(\rho)=1-\frac{2M}{r(\rho)}+\frac{e^2}{r^2(\rho)}\left(1+\frac{a^2}{5r^2(\rho)}\right)
\end{equation}
and
\begin{equation}
f(\rho)=\frac{1}{F(\rho)}\left(1-\frac{e^2a^2}{5r^4(\rho)}\right).
\end{equation}
We now seek location of mean metric horizons.
\section{Horizons location for mean metric}
One can obtain event horizon location of the mean metric (15) by
solving $F(\rho_{EH})=0$ and location of apparent (interior and
exterior) horizons by solving null condition
$g^{\mu\nu}\partial_{\mu}\rho\partial_{\nu}\rho=0$
 which leads to the equation
$F(\rho_{AH})=0$ such that
\begin{equation}
1-\frac{2M}{r_H}+\frac{e^2}{r_H^2}+\frac{e^2a^2}{5r_H^4}=0.
\end{equation}
The above equation has not exactly analytic solution for $a\neq0$
but for small $a$ we can use perturbation series expansion to
evaluate the event horizon location. To do so we first define
$\epsilon=\frac{a}{r_H}$ for which the horizon equation (18) can
be written as $r_H^2-2Mr_H+e^2(1+\epsilon^2)=0.$ The latter
equation has a real solution as $
r_H=M\{1+\sqrt{1-(e^2/M^2)(1+\epsilon^2/5)}\}$ for
$\frac{e^2}{M^2}(1+\frac{\epsilon^2}{5})<1.$ We know that for a
single RN black hole $\frac{e}{M}<1$ and so the  condition
$\frac{e^2}{M^2}(1+\frac{\epsilon^2}{5})<1$  reads $
\epsilon<1(a<r_H)$ for which horizon of the ensemble of
statistical RN black holes  does not destructed by raising
$0<\epsilon<1$ if we want to apply perturbation series expansion
method to obtain asymptotically behavior of the event horizon
solution versus the parameters $(a, e, M).$ Thus we must be obtain
perturbation series expansion form of the event horizon but for
$a<r_H$ as follows. Inserting
 \begin{equation}
r^{\pm}_{H}=r^{\pm}_0+ar^{\pm}_1+a^2r^{\pm}_2+O(a^3)\end{equation}
 and solving (18) as order by order we obtain \begin{equation}
 r^{\pm}_0=M\pm\sqrt{M^2-e^2},~~~r^{\pm}_1=0,~~~r^{\pm}_2=\frac{\mp e^2}{10\sqrt{M^2-e^2}(M\pm\sqrt{M^2-e^2})^2}\end{equation} where $r_H^+$
 and $r_H^-$  denote to apparent exterior  and Cauchy (interior) horizon radiuses of the mean metric (7) respectively.
Inserting (9) and
 (19) one can obtain perturbation series expansion of the equation (12) which up to terms in order of $O(a^3)$ become: \begin{equation} \rho_H=\rho_0^{\pm}+a^2 \rho_2^{\pm} \end{equation}
 where we defined \begin{equation} \rho_0^{\pm}=r^{\pm}_0,~~~\rho_2^{\pm}=r^{\pm}_2+\frac{2Mr^{\pm}_0-e^2}{10{r^{\pm}_0}^3}.\end{equation}
Area equation of apparent horizon hypersurface of the spherically
symmetric static
  mean metric (15) is defined by  $A=4\pi \rho^2_{H}$ which up to terms in order of $O(a^3)$
  reads \begin{equation} A^{\pm}=A^{\pm}_0+a^2A^{\pm}_2\end{equation}
   where we defined
  \begin{equation} A^{\pm}_0=4\pi (r_0^{\pm})^2,~~~A_2^{\pm}=8\pi\bigg[r_0^{\pm}r_2^{\pm}+\frac{1}{10
  r_0^{\pm}}\bigg(2M-\frac{e^2}{r_0^{\pm}}\bigg)\bigg].\end{equation}
 According to Bekenstein-Hawking entropy theorem we can result,
$A_+(A_-)$ given by (23) will be entropy function of exterior
(interior) horizon of the mean metric (15).  Black holes
containing multiple horizons have corresponding several
temperatures. Such a black hole will be in-equilibrium thermally
throughout the space time where the temperature has a gradient
between the horizons. Thermal equilibrium is possible only if
horizon radiuses and so the corresponding temperatures become
equal (see for instance [22, 23]). The latter situations are
happened for an extreme RNBH where $M=e$ and so $r_{H}^+=r_H^-.$
We now calculate thermodynamic characteristics of interior and
exterior horizons of the non-extreme mean metric of RNBHs
statistical ensemble.
\section{Mean metric thermodynamics}
 In the next section we will
consider massless, charge-less quantum scalar field effects on
luminosity of the quantum perturbed coarse graining RNBHs where
its electric charge become invariant quantity. Hence it is useful
to define dimensionless black hole mass $m=\frac{M}{e}$ and
ensemble factor $\delta=\frac{a}{e}$ in what follows.  In the
latter case exterior horizon entropy of mean metric (15) can be
obtained up to terms in order of $O(\delta^3)$ as follows.
\begin{equation}
S_+(m,\delta)=(m+\sqrt{m^2-1})^2+\frac{\delta^2}{5}\bigg[\frac{2(m^2-1)^\frac{3}{2}+m(2m^2-3)}{\sqrt{m^2-1}(m+\sqrt{m^2-1})^2}\bigg]\end{equation}
and its interior horizon entropy become
\begin{equation} S_-(m,\delta)=(m-\sqrt{m^2-1})^2+\frac{\delta^2}{5}\bigg[\frac{2(m^2-1)^\frac{3}{2}-m(2m^2-3)}{\sqrt{m^2-1}(m-\sqrt{m^2-1})^2}\bigg]\end{equation}
where   $0<\delta<1$ and \begin{equation}
S_{\pm}=\frac{A_{\pm}}{4\pi e^2}>0.\end{equation} Diagrams of the
entropies (25) and (26) are plotted versus  $m$ in figure 4. They
show that $S_{\pm}>0$ for a single RNBH ($\delta=0$) in limits
$m\to1$ but for
 an ensemble of RNBHs for which we use $\delta=0.9,$ they reach to infinity $S_{\pm}\to\mp\infty.$ In fact
for physical systems the entropy itself must be positive function
but its
 variations may to be reach to some negative values. Hence we define difference between interior horizon entropy and exterior horizon entropy
 as
\begin{equation} \Delta S=S_+-S_-=4m\sqrt{m^2-1}+\frac{2\delta^2}{5}\bigg[\frac{m(2m^2-1)(2m^2-3)}{\sqrt{m^2-1}}-4m(m^2-1)^\frac{3}{2}\bigg]
\end{equation}
and total entropy such as follows.
\begin{equation}S_{tot}=S_++S_-=4m^2-2+\frac{4\delta^2}{5}(4m^4-6m^2+1).\end{equation} Diagrams of $\Delta S$ and $S_{tot}$ are plotted in figure 3.
 Fortunately these diagrams show that for a single RNBH where $\delta=0,$ we will have
 $\Delta S>0$ by decreasing $m\to1$ and $S_{tot}>0$ but for ensemble of RNBHs
with $\delta=0.9$ we have $\Delta S<0$ while $S_{tot}>0$. Hence
$\Delta S$ and $S_{tot}$ should be considered as physical
entropies of coarse graining RNBHs.   Decrease of entropy causes
to some negative temperatures (see figure 2) in thermodynamic
systems containing bounded energy levels. In the latter case there
is a critical temperature for which the system exhibits with a
phase transition reaching to Bose-Einstein condensation state
microscopically. In thermodynamics, increase of entropy $\Delta
S>0$ means an increase of disorder or randomness in natural
systems. It measures heat transfer of the system for which heat
flows naturally from a warmer to a cooler substance. Decrease of
entropy $\Delta S<0$ means an increase of orderliness or
organization of microstates of a system. To do so the substance of
a system must be lose heat in the transfer process. Individual
systems can experience negative entropy, but overall, natural
processes in the universe trend toward positive entropy. Negative
entropy was first introduced for living things by
 Ervin Schr\"{o}dinger in 1944 as the reverse concept of entropy, to
describe the order that can emerge from chaos [24]. The heat
generated by computations in the information theory is other
applications for negative entropy concept (see [25-28] for more
discussions). However we consider $\Delta S$ and $S_{tot}$ to be
physical entropies of RNBHs statistical ensemble containing two
horizons which is in accord to positivity condition of the
Bekenstein-Hawking entropy theorem. Our coarse graining RNBHs can
be considered as a two level thermodynamical system with upper
bound finite energy $M$ because it has two dual (interior and
exterior) horizons. We now calculate exterior (interior) horizon
temperature $T_{+}(T_{-})$ of the RNBHs mean metric (15) as
follows.
\begin{equation} T^*_{\pm}=(4\pi
e)T_{\pm}=\frac{1}{\big(\frac{\partial S_{\pm}}{\partial
m}\big)_\delta}=\pm\frac{1}{2}\frac{\sqrt{m^2-1}}{(m\pm\sqrt{m^2-1})^2}$$$$
+\frac{\delta^2}{60}\bigg[\frac{4m-2m^3-2m^5\mp\sqrt{m^2-1}(2m^4+3m^2-3)}{(m^2-1)^\frac{5}{6}(m\pm\sqrt{m^2-1})^6}\bigg]\end{equation}
Their diagrams are plotted against $m$ in figure 2 for $\delta=0;
0.9$. For $m>>1$ we see that $T^*_-(T^*_+)$ has some negative
(positive) values and their sign is changed when $m\to1$. We also
plotted diagram for $T^*_\pm$ versus $\Delta T^*=T^*_+-T^*_-$ in
figure 2. They show that $T^*_-<0$ for $\Delta T^*>0$ reaching to
zero value at $\Delta T^*=0$ for $\delta=0,0.9.$ While
$T_+^*>0(T_+^*<0)$ when $\Delta T^*\to0^+$ for $\delta=0(0.9)$
after than to obtain a finite  positive maximum  value. This
maximum has smaller value for $\delta=0.9$ with respect to
situations where we choose $\delta=0.$
 In ordinary statistical physics, negative
temperatures are taken into account when the system has  upper
bound (maximum finite) energy for which entropy is continuously
increasing but the energy and temperature decrease and vice versa.
In the latter case the system
 reaches to Bose-Einstein condensation state  microscopically. Energy upper bound of our system is its total mass $M$ for which we have $m>1.$
 Regarding quantum matter effects on mean metric we will show in section
 5,
mass of mean metric decreases finally as $m_{final}=1$ (see
figures 1). Bose-Einstein condensation state needs a phase
transition which is happened when sign of heat capacity is changed
. Hence we now calculate interior and exterior horizon of mean
metric heat capacity $C_{\pm}^*$ which up to terms in order of
$O(\delta^3),$
 at constant electric charge $e$ and
ensemble radius $a$ become
\begin{equation}{C_{\pm}}^* =\frac{C_{\delta}^{\pm}}{4\pi e^2}=\bigg(T_{\pm}\frac{\partial S_{\pm}}{\partial T_{\pm}}\bigg)_{\delta}=
\bigg(\frac{\partial T^*_{\pm}}{\partial
m}\bigg)^{-1}_{\delta}=-\frac{2\sqrt{m^2-1}(m\pm\sqrt{m^2-1})^2}{2\sqrt{m^2-1}\mp
m}$$$$
+\frac{2\delta^2}{45}\bigg[\frac{2m\sqrt{m^2-1}(4m^4+12m^2-15)\pm8m^6\pm20m^4\mp49m^2\pm21}{(m^2-2\pm
m\sqrt{m^2-1})^2(m^2-1)^{\frac{5}{6}}}\bigg].\end{equation} Their
diagrams are plotted against $m$ in figure 5. They show that sign
of  $C_+^*$ is changed at $m_c=1.15(1.2)$ for $\delta=0(0.9)$ but
sign of $C_-^*$ is changed at $m=1$ for $\delta=0,0.9.$ We plot
also diagrams of $C_{\pm}^*$ versus $\Delta T^*$ in figure 5. They
show a changing of sign for $C_+^*$ when $\Delta T^*\to0$ and
$\delta=0,0.9$ but not for $C_-^*.$ In case $\delta=0.9$ we see
$C_-^*<0$ for $\Delta T^*>0$ but its absolute value exhibits with
a minimum value. When $\Delta T^*\to0$ we see $C_-^*$ which
decreases monotonically to negative infinite value for $\delta=0.$
Changing of sign of exterior horizon heat capacity means that
there is happened a phase transition when the quantum perturbed
RNBHs ensemble reaches to its stable state with minimum mass
$m_{final}=1.$ To determine order kind of this phase transition we
should  study behavior of the corresponding Gibbs free energy
as follows.\\
Exterior and interior horizon Gibbs free energies are defined by
\begin{equation}G_{\pm}=M-T_{\pm}A_{\pm}-\Phi_{\pm}\end{equation}
where entropy $A_{\pm}$ is given by (24) and electric potential
$\Phi_{\pm}$ is defined by
\begin{equation}\Phi_{\pm}=-T_{\pm}\bigg(\frac{\partial A_{\pm}}{\partial e}\bigg)_{a,M}.\end{equation}
Inserting $M=em, A_{\pm}=4\pi e^2 S_{\pm}$ and the equation (30)
the above Gibbs energy equation reads
\begin{equation}G_{\pm}^* =\frac{G_{\pm}}{e}=m-T^*_{\pm}S_{\pm}-8\pi S_{\pm}-4\pi e\frac{\partial S_{\pm}}{\partial e}\end{equation}
in which we have
\begin{equation}e\frac{\partial S_{\pm}}{\partial e}=\mp\frac{2m(m\pm\sqrt{m^2-1})^2}{\sqrt{m^2-1}}-\frac{\delta^2}{15(m^2-1)^\frac{11}{6}
(m\pm\sqrt{m^2-1})^2}$$$$\times
[28m^7-75m^5+74m^3-27m\pm\sqrt{m^2-1}(26m^6-61m^4+47m^2-12)].\end{equation}
We plot diagrams of the above equations against $m$ in figure 6.
They show that $G_-^*$ has minimum zero value at $m=1$ but $G_+^*$
raises to $+\infty$ by decreasing $m\to1$ for $\delta=0.9.$ In
case $\delta=0,$ we see $G_{\pm}^*\to\pm\infty$ when $m\to1.$
Furthermore we plot diagrams of $G_{\pm}^*$ versus $\Delta T^*$ in
figure 6. We see  $G_-^*\to-\infty$ when $\Delta T^*\to0$ for
$\delta=0$ but $G_-^*\to0^+$ for $\delta=0.$ $G_{+}^*$ decreases
to a positive minimum value by decreasing $\Delta T^*\to0$ and
then reaches to positive infinite value. The latter behavior shows
changing the sign of first derivative of $G_+^*$ when decreases
$m$ and/or $\Delta T^*$ which means that the
phase transition is first order.\\
One of other suitable quantities which should be calculated is
pressure of black hole micro-particles which coincide with the
interior horizon as follows. If a quantum particle is collapsed
inside of the interior (exterior) horizon then its de Broglie wave
length must be at least $\lambda_-\approx
2\rho_-(\lambda_+\approx2\rho_+).$ We use de Broglie quantization
condition on quantum particles as
$p_{\pm}=\frac{h}{\lambda_{\pm}}$ where $h$ is Planck constant and
$p_{\pm}$ is momentum of in-falling quantum particles inside of
the horizons. In Plank units where $c=h=G=1$ we can write
\begin{equation} \Delta
p=p_--p_+=\frac{1}{2}\bigg(\frac{1}{\rho_-}-\frac{1}{\rho_+}\bigg)\end{equation}
in which $\Delta p$ is difference of momentum of quantum particles
which  move from exterior horizon $\rho_+$ to interior $\rho_-$
horizon. For $c=1$ they move for durations $\Delta
t=\rho_+-\rho_-.$ We now use the latter assumptions to rewrite
Newton`s second law as
\begin{equation} F=\frac{\Delta p}{\Delta
t}=\frac{1}{2\rho_+\rho_-}. \end{equation} $F$ is dimensionless
force which affects on interior horizon surface. When the system
become stable mechanically then $F$ must be balanced by the
electric force of the system defined by
$F_{E}=e\big(\frac{\Phi_--\Phi_+}{\rho_+-\rho_-}\big).$
Spherically symmetric condition of the system causes to choose
some radial motions for quantum particles located inside of the
statistical ensemble of RNBHs. However one can define pressure of
moving charged quantum particle on the interior horizon as
\begin{equation} P_-=\frac{F}{4\pi \rho^2_-}=\frac{1}{8\pi
\rho_+\rho_-^3}\end{equation} which by inserting (22) and using
some simple calculations reads
\begin{equation}P^*_-(m)=8\pi e^4P_-=(m-\sqrt{m^2-1})^{-2}\bigg\{1+\frac{\delta^2}{5}$$$$\times\bigg[\frac{24m^5-26m^3+2m^2-6m-2}{(m-\sqrt{m^2-1})^3}
$$$$-\frac{(24m^6-38m^4+2m^3+16m^2-3m-2)}{\sqrt{m^2-1}(m-\sqrt{m^2-1})^3}\bigg]\bigg\}.\end{equation}
 We plot diagram of the above pressure in figure 8. They show that $P^*_->0(<0)$ in case $\delta=0(0.9)$ for all values
 of $\Delta T^*>0.$ Diagrams show that $P_-^*$ is
 vanishing when $\Delta T^*\to0.$ Also we plot diagram for $P_-^*$
 versus $m.$ It shows $P_-^*\to0^+$ for $\delta=0.$ In case where $\delta=0.9$ one can see $P_-^*<0$ when $m\to1$ for $m>1$ but $P_-^*\to+\infty.$
  The latter results predict dark matter behavior of the interior horizon matter counterpart where for positive mass $m>1$
 there is some `negative` pressure. How can decreases mass of mean metric
 RNBHs?  Dynamically this is possible if we consider corrections of
 quantum matter field interacting with the mean metric of RNBHs as follows.
 This makes as unstable quantum mechanically the mean metric of RNBHs.
In the next section we assume interaction of the mean metric of
RNBHs statistical ensemble with mass-less, charge-less quantum
scalar field for which $e$ will be invariant of the system and so
there is not any electromagnetic radiation. In other words there
will be only mass interaction between quantum scalar field and
ensemble of the RNBHs. They reduce usually to the well known
Hawking thermal radiation of the quantum perturbed mean metric
which is causing to mass-loss of the mean RNBHs. For such a
quantum mechanically unstable mean
 metric we now calculate its luminosity, mass loss process and switching off effect.
\section{Mean quantum RNBH mass loss}
 We applied massless, charge-less quantum scalar field
Hawking thermal radiation effects on single quantum unstable RNBH
and calculated time dependence mass loss function in ref. [2]. We
obtained that the evaporating quantum perturbed RNBH exhibits with
switching off effect before than that its mass disappear
completely. It should be pointed that electric charge of the black
hole is invariant of the system because there is no
electromagnetic interaction between its electric charge and
charge-less quantum matter scalar field. Thus  mass of
 the RNBH decreases to reach to non-vanishing remnant stable mini
Lukewarm black hole with $m_{final}=1$. In other words its
luminosity is eliminated while its mass does not eliminated
completely (see figures 9, 10 and 11 given in ref. [2]). Here we
study mass loss and switching off effect of quantum perturbed mean
metric (15). This is a dynamical approach to describe that how
mean metric of RNBHs statistical  ensemble exhibits with a phase
transition leading to a possible Bose-Einstein condensation state
microscopically. Line element of the evaporating mean metric (15)
can be written near the exterior horizon as Vaida form (see for
instance [29]):
\begin{equation}
ds^2\simeq\bigg(1-\frac{r_+(v)}{r}\bigg)dv^2-2dvdr-r^2(d\theta^2+\sin^2\theta
d\varphi^2)\end{equation} with the associated stress energy tensor
\begin{equation} <\hat{T}_{\mu\nu}^{quant}>_{ren}=\frac{1}{4\pi
r^2}\frac{dr_+(v)}{dv}\delta_{\mu v}\delta_{\nu v}\end{equation}
where $(v,r)$ is advance Eddington-Finkelstein coordinates system.
Subscript $ren$ denotes to the word $Renormalization$, and $<>$
denotes to expectation value of quantum matter scalar field stress
tensor operator evaluated in its vacuum state.   The black hole
luminosity is defined by the following equation from point of view
of distant observer located in $r.$
\begin{equation} L(r,v)=4\pi r^2 <\hat{T}^r_v>^{quant}_{ren}.\end{equation}
Applying (40) and (41) the equation (42) become
\begin{equation} L=-\frac{1}{2}\frac{dr_+(v)}{dv}\end{equation} where
negative sign describes inward flux of negative energy across the
horizon. This causes to shrink the mean metric horizon of RNBHs
statistical ensemble. In the latter case quantum particles of
matter content of the black hole are in high energy state and so
one can assume that the quantum black hole behaves as a black body
radiation  which its luminosity is defined by well known
Stefan-Boltzman law as follows.
\begin{equation} L=\sigma_{SB} A T^4\end{equation} where $A$
is surface area of the black body, $T$ is its temperature.
$\sigma_{SB}=5.67\times10^{-8}\frac{J}{m^2\times{^{\circ}K}^{4}
\times Sec}$ is Stefan-Boltzman coupling constant which its
dimensions become as $(length)^2$  in units $G=c=1$. If (44)
satisfies (43), then we can obtain mass loss equation of the mean
metric of RNBHs statistical ensemble such that
\begin{equation} \frac{dr_+(v)}{dv}=-2\sigma_{SB}\xi
A_+(v)T^4_+(v)\end{equation} where the normalization constant
$\xi$ depends linearity on the number of massless charge-less
quantum matter fields and will control the rate of evaporation.
Inserting (27) one can show that the luminosity (44) for RNBHs
mean metric (15) become
\begin{equation}  L_+^*(m)=\frac{(4\pi)^3e^2}{\sigma_{SB}}L=S_+(m){T^*}^4_{+}(m)\end{equation}
where $S_+(m)$ and $T_+^*(m)$ should be inserted from the
equations (25) and (30) respectively.  Applying (19), (20), (21),
(22), (27) and some simple calculations we can show that the mean
mass-loss equation (45) for RNBHs mean metric (15) reads
\begin{equation} \Delta v^*=v^*(m)-v^*_{\infty}=-\frac{1}{2}\int_m^1\bigg(\frac{dr_+}{d\rho_+}\bigg)\frac{dm}{S_+^\frac{3}{2}{T^*_+}^5}\end{equation}
where we used (19), (20), (21), (22), (27), $\delta=\frac{a}{e},$
$m=\frac{M}{e},$ and $\rho_+=e\sqrt{S_+}$ to calculate
$\frac{dr_+}{d\rho_+}$ which up to terms in order of $O(\delta^3)$
become
\begin{equation}\frac{dr_+}{d\rho_+}=1+\frac{\delta^2}{5}\bigg(\frac{2}{(m+\sqrt{m^2-1})^3}-\frac{3}{(m+\sqrt{m^2-1})^4}\bigg).\end{equation}
$v^*_{\infty}=v^*(1)$ given in the equation (47) is integral
constant for which evaporating mean mass of RNBHs statistical
ensemble reaches to its final value as $m_{final}=1.$ Also we
defined dimensionless advance Eddington-Finkelstein time
coordinate $v^*$ as follows.
\begin{equation}\frac{v^*}{v}=\frac{2\xi \sigma_{SB}}{(4\pi
e)^3}.\end{equation} When exterior horizon  of quantum evaporating
RNBHs mean metric reduces to scale of its interior horizon as
$r_+(v)\to r_-$ then one can use similar equations for luminosity
and mass-loss equation (46) and (47) for interior horizon as
follows.
\begin{equation}L_-^*(m)=\frac{(4\pi)^3e^2}{\sigma_{SB}}L=S_-(m){T^*}^4_{-}(m)\end{equation}
\begin{equation} \Delta v^*=v^*(m)-v^*_{\infty}=-\frac{1}{2}\int_m^1\bigg(\frac{dr_-}{d\rho_-}\bigg)\bigg(\frac{dS_
-}{dm}\bigg)\frac{dm}{S_-^\frac{3}{2}{T^*_-}^4}
\end{equation}
where
\begin{equation}\frac{dr_-}{d\rho_-}=1+\frac{\delta^2}{5}\bigg(\frac{2}{(m-\sqrt{m^2-1})^3}-\frac{3}{(m-\sqrt{m^2-1})^4}\bigg).\end{equation}
 Diagrams of the luminosity (46),(50) and the evaporating mean RNBHs mass loss equation (47), (51) are plotted versus mass
parameter $m$ in figure 7. They show that evaporating quantum
unstable mean mass of RNBHs final state reaches to remnant stable
cold mini Lukewarm RNBH with final mass $m_{final}=1$ where its
causal singularity is still covered by its shrunken horizon and
its luminosity reaches to zero value. We see that invariant
conditions on the black hole electric charge $e$ causes to valid
the Penrose cosmic censorship hypothesis while the black hole
metric is evaporated where the  casual singularity of mean metric
(15) defined by $\rho=0$ is still covered by their smallest scale
horizons hyper-surface with no naked singularity.
\section{ Summary and Discussion} According to the Debbasch
approach we calculated mean metric of RNBHs statistical ensemble
to obtain locations of interior and exterior horizons. We
calculated corresponding entropy, temperature, heat capacity,
Gibbs free energy and pressure. At last section of the paper we
considered interaction of massless, charge-less quantum scalar
matter field on quantum perturbed mean metric of  coarse graining
RNBHs. Our mathematical calculations predict evaporation of the
mean metric which reduces to a remnant stable mini black hole
metric with non-vanishing mass. Before than the evaporation
reaches to its final state the mean metric exhibits with a first
order phase transition and there is happened Bose-Einstein
condensation state microscopically. Our results approve outputs of
the published work [2] qualitatively in which the author studied
thermodynamic
behavior of a single RN black hole.\\

\begin{figure}[tbp] \centering
\includegraphics[width=5cm,height=5cm]{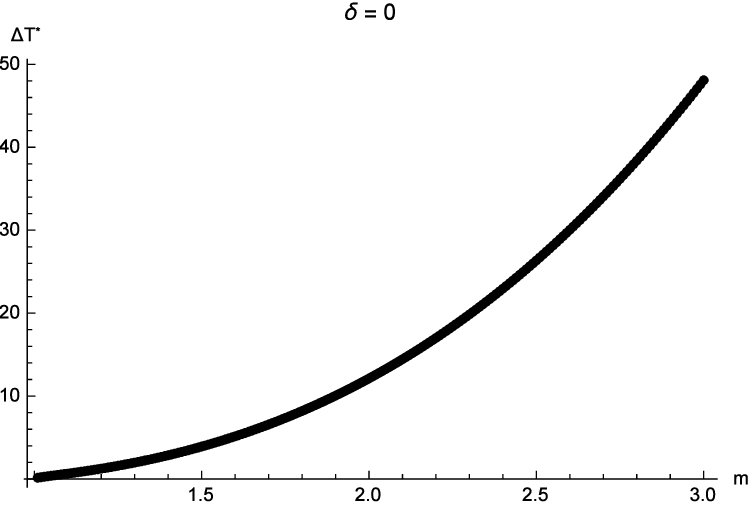}
\includegraphics[width=5cm,height=5cm]{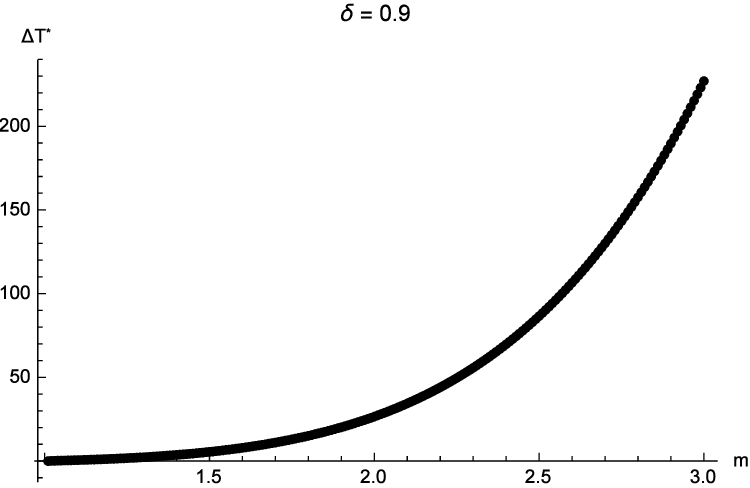}
\includegraphics[width=5cm,height=5cm]{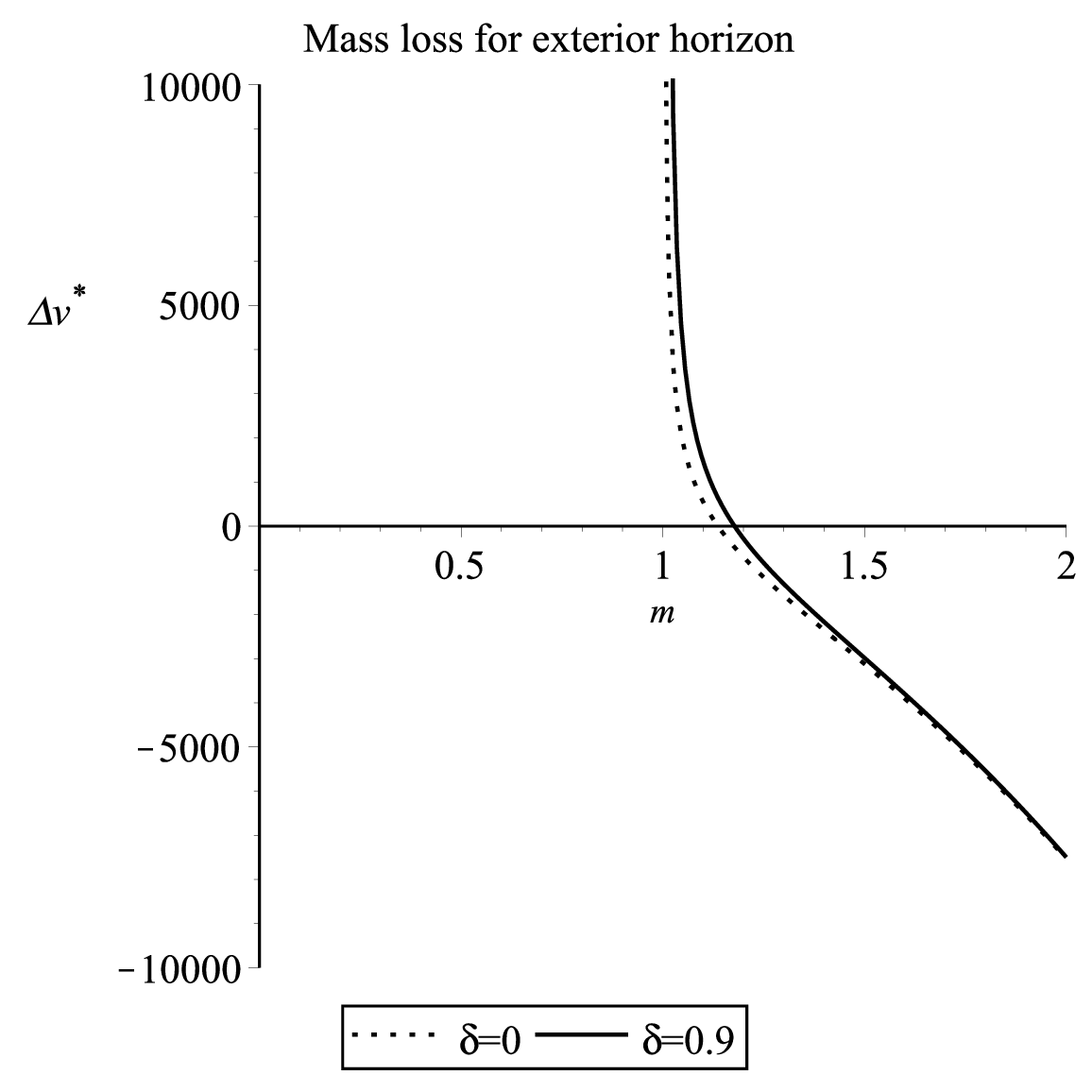}
   \includegraphics[width=5cm,height=5cm]{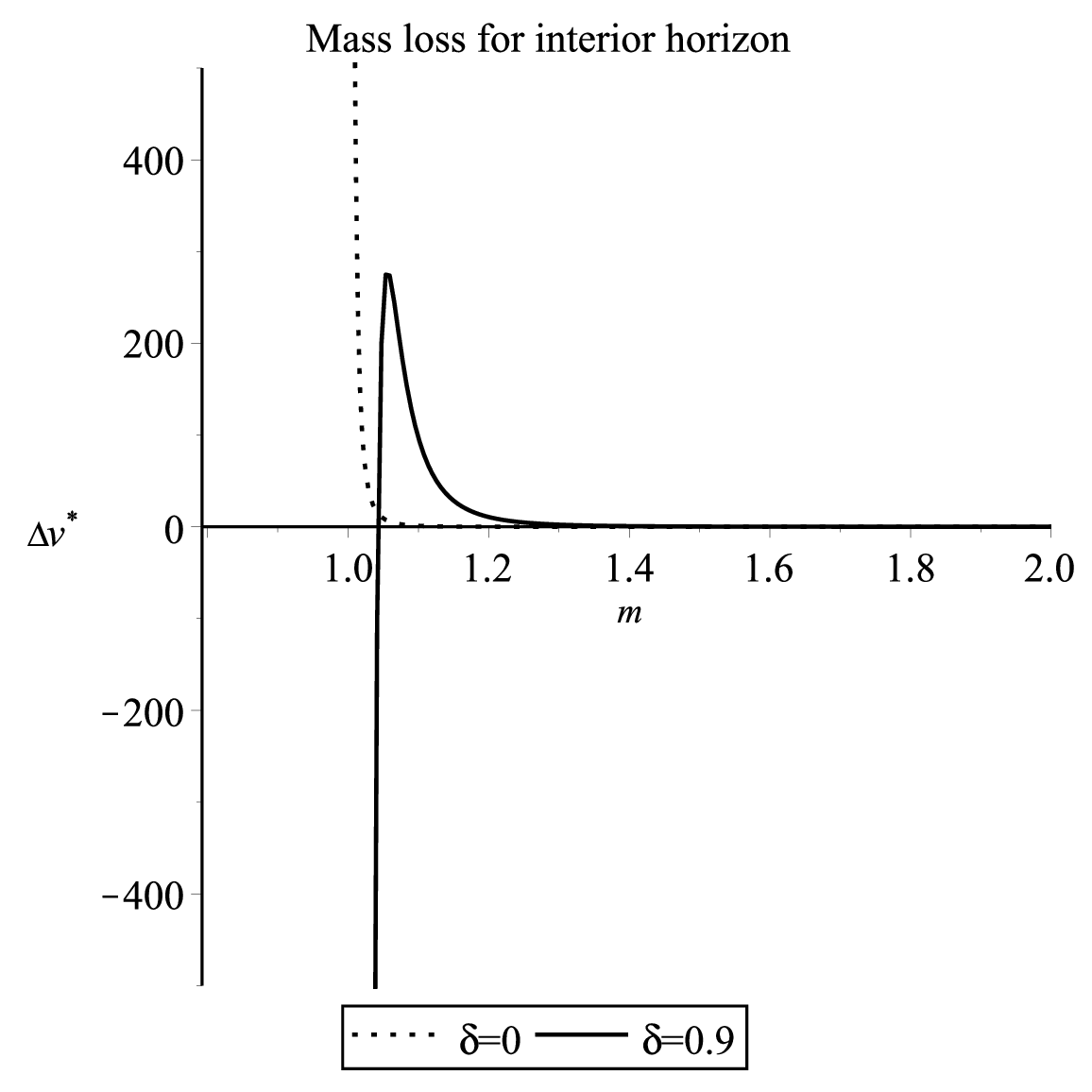}
   \includegraphics[width=5cm,height=5cm]{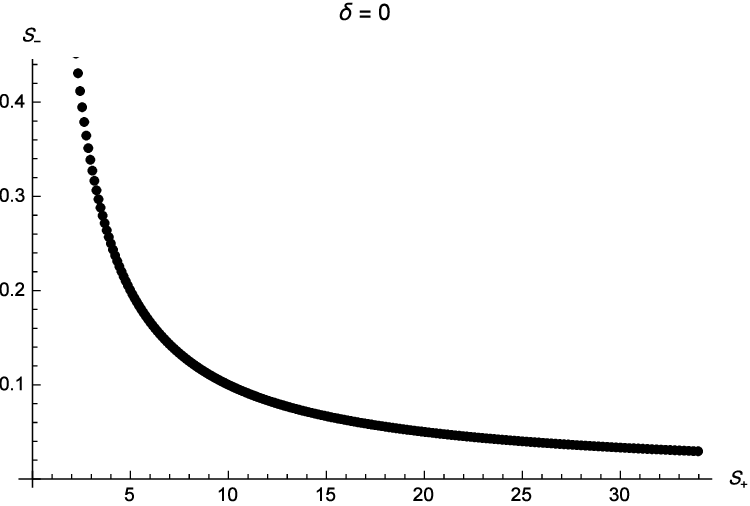}
    \includegraphics[width=5cm,height=5cm]{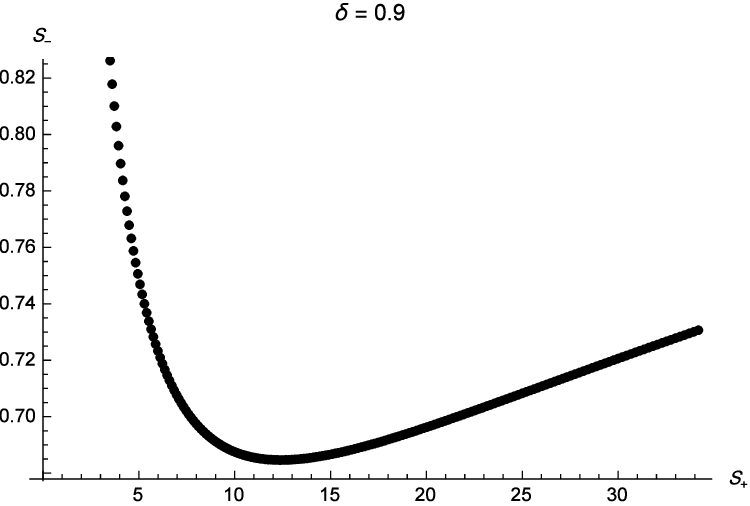}
\caption{\label{fig:epsart} Diagram of mass-loss $m(v)$, interior
(exterior) horizon entropy $S_-(S_+)$ and difference of
temperatures between interior and exterior horizons $\Delta
T^*(m)$ are plotted against for single RNBH $\delta=0$ and mean
metric of coarse graining RNBHs $\delta=0.9$. }
\end{figure}
\begin{figure}[tbp] \centering
\includegraphics[width=6cm,height=6cm]{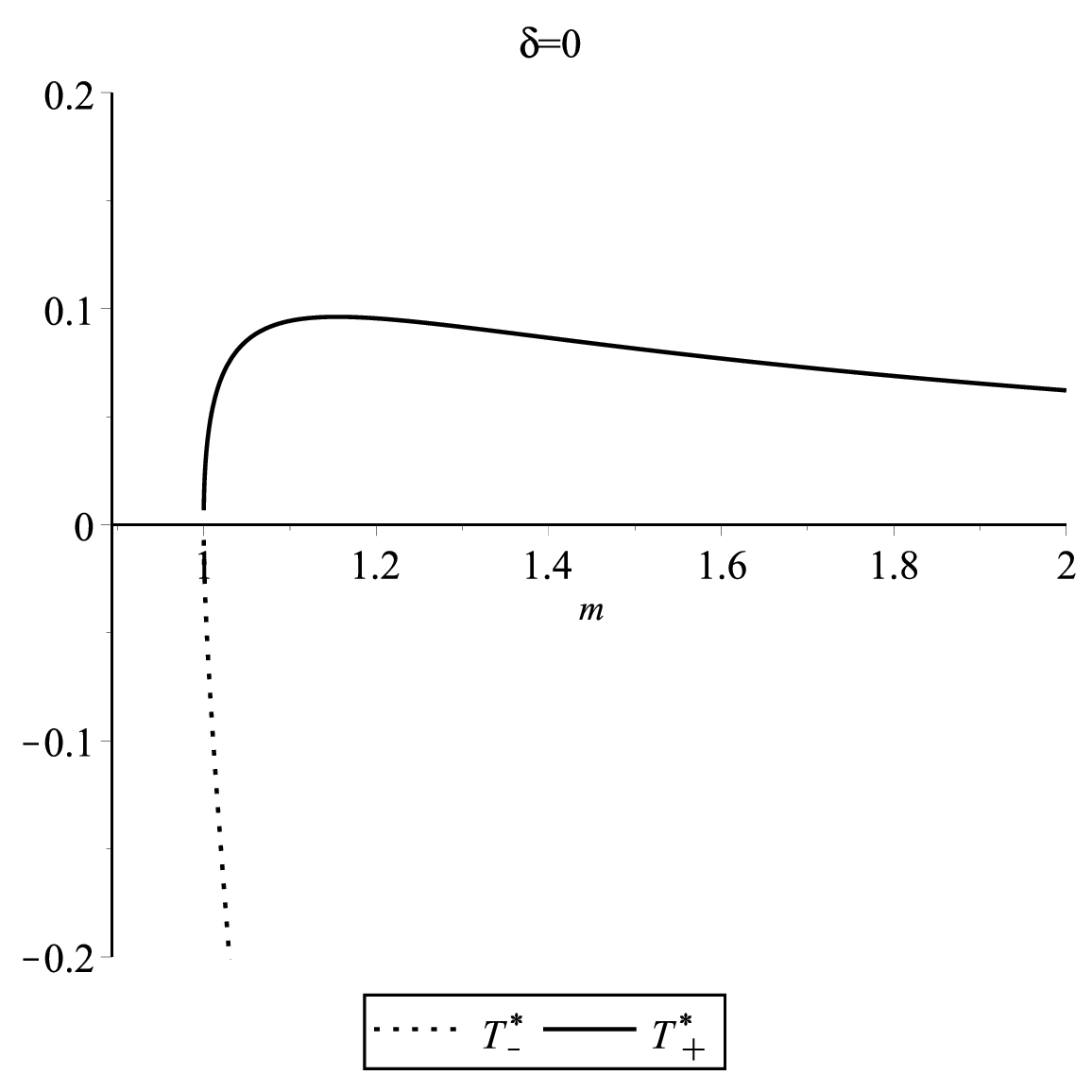}
     \includegraphics[width=6cm,height=6cm]{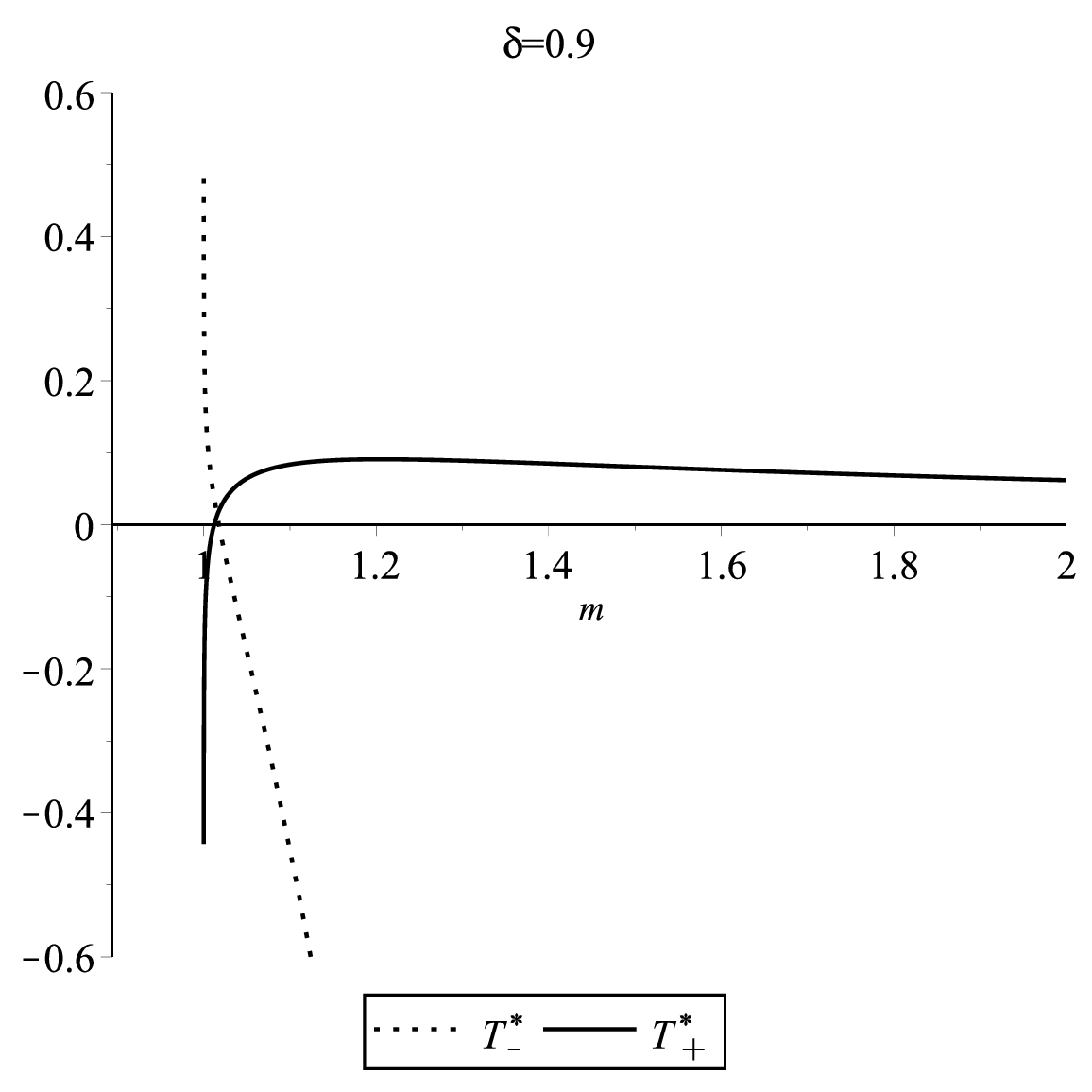}
\includegraphics[width=6cm,height=6cm]{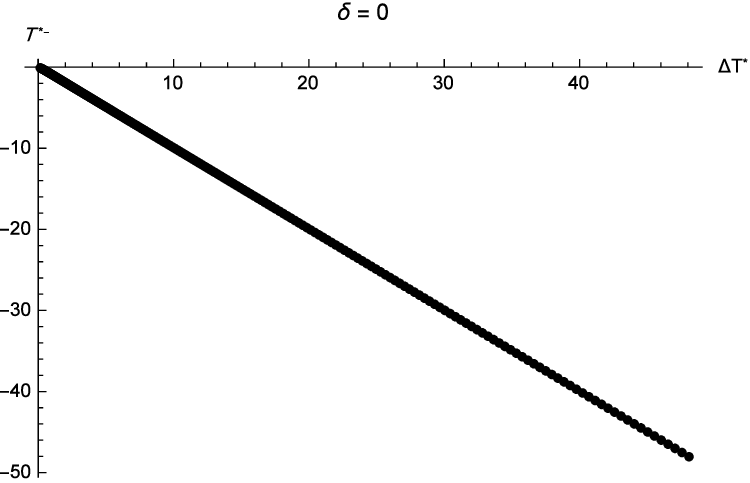}
\includegraphics[width=6cm,height=6cm]{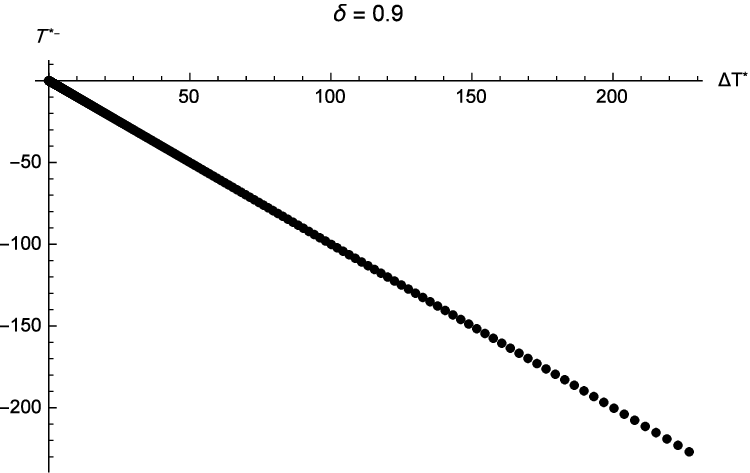}
\includegraphics[width=6cm,height=6cm]{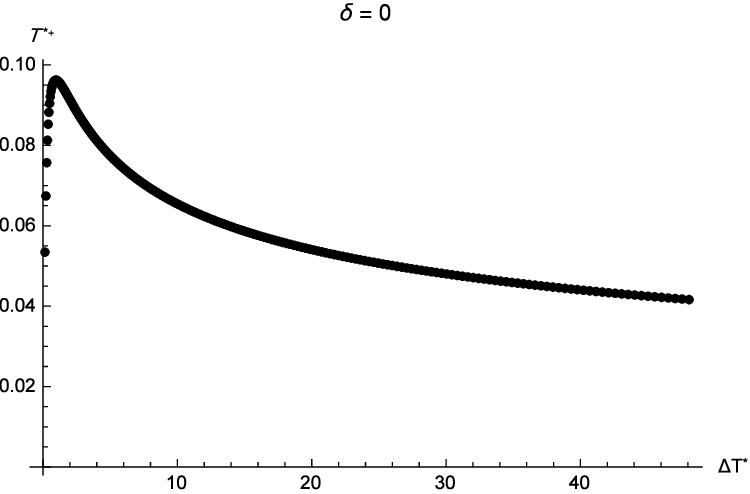}
\includegraphics[width=6cm,height=6cm]{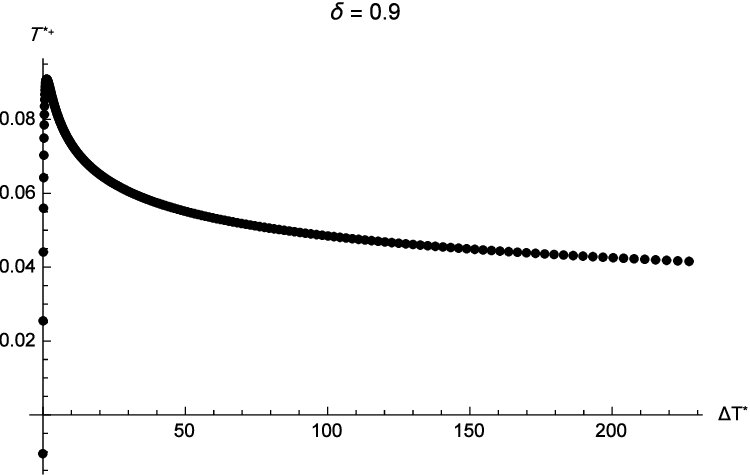}
\caption{\label{fig:epsart} Diagrams of $T_{\pm}^*$ are plotted
against $m$ and $\Delta T^*$ for $\delta=0$ and $\delta=0.9$. }
\end{figure}
\begin{figure}[tbp] \centering
\includegraphics[width=6cm,height=6cm]{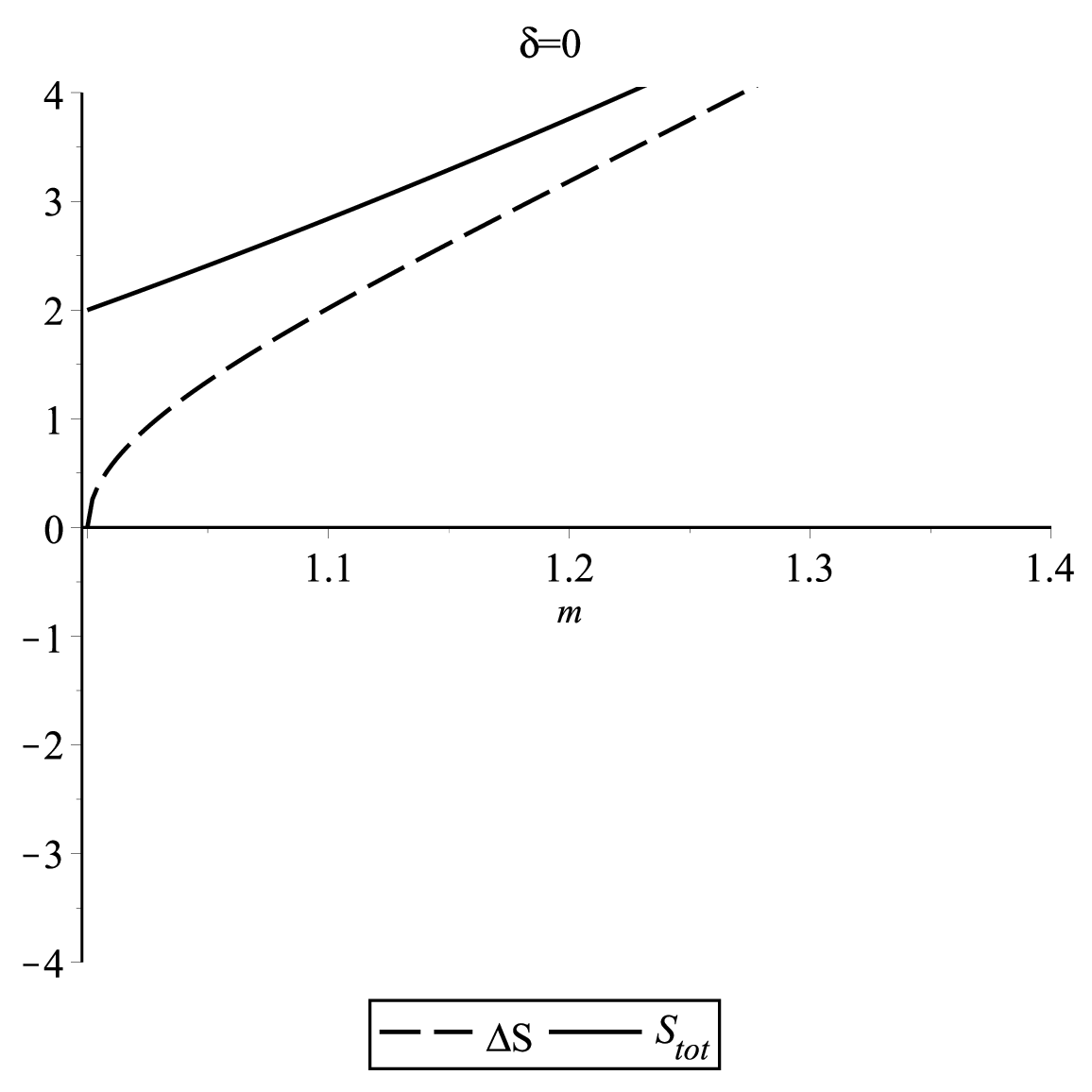}
\includegraphics[width=6cm,height=6cm]{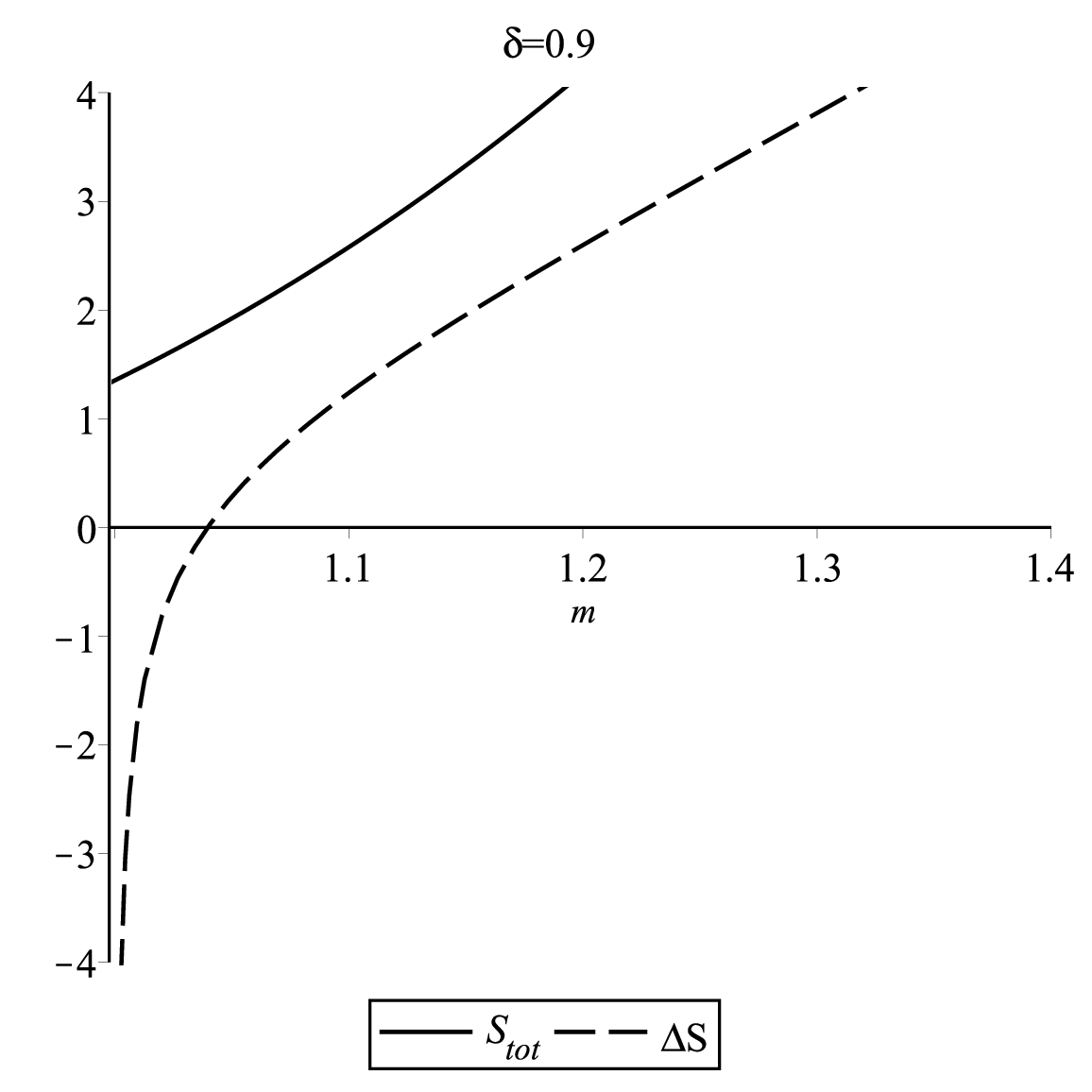}
\includegraphics[width=6cm,height=6cm]{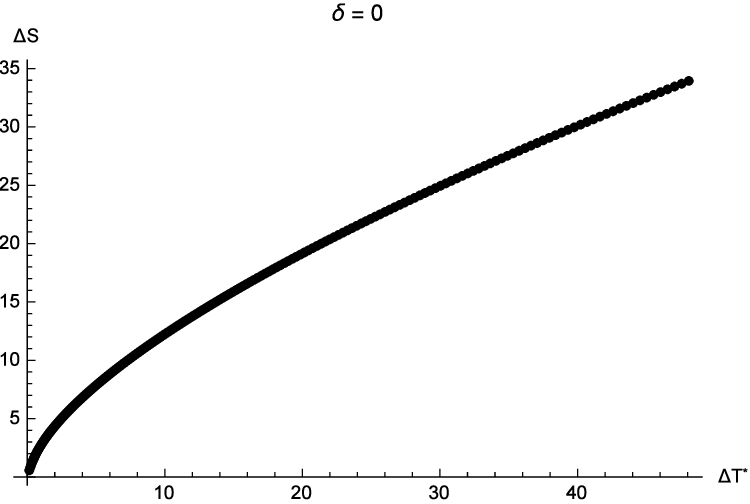}
 \includegraphics[width=6cm,height=6cm]{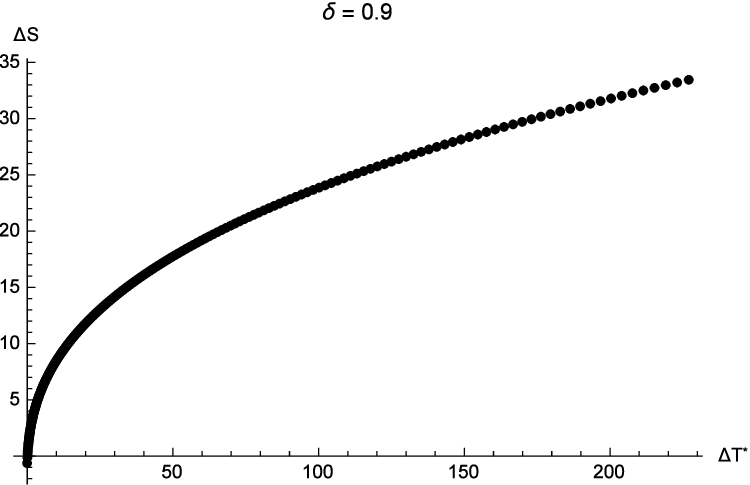}
 \includegraphics[width=6cm,height=6cm]{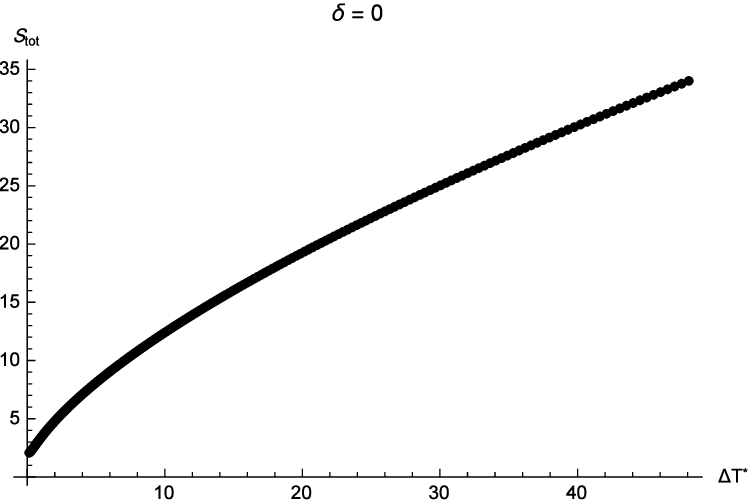}
 \includegraphics[width=6cm,height=6cm]{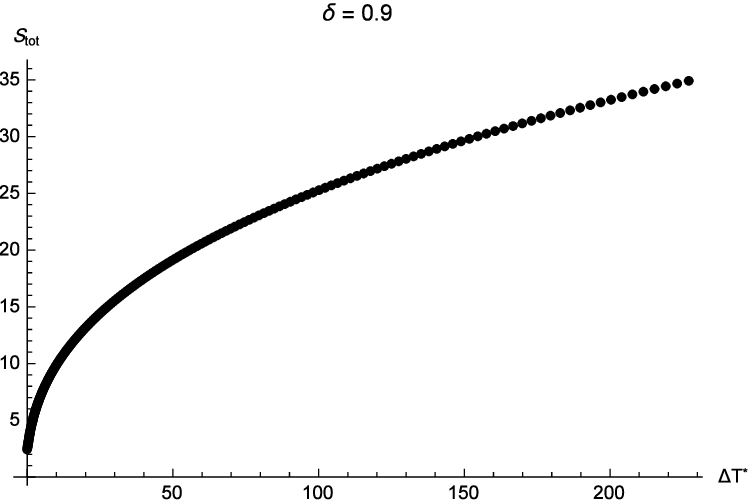}
\caption{\label{fig:epsart} Diagrams of $\Delta S$ and $S_{tot}$
are plotted against $m$ and $\Delta T^*$ for $\delta=0$ and
$\delta=0.9$. }
\end{figure}
\begin{figure}[tbp] \centering
 \includegraphics[width=6cm,height=6cm]{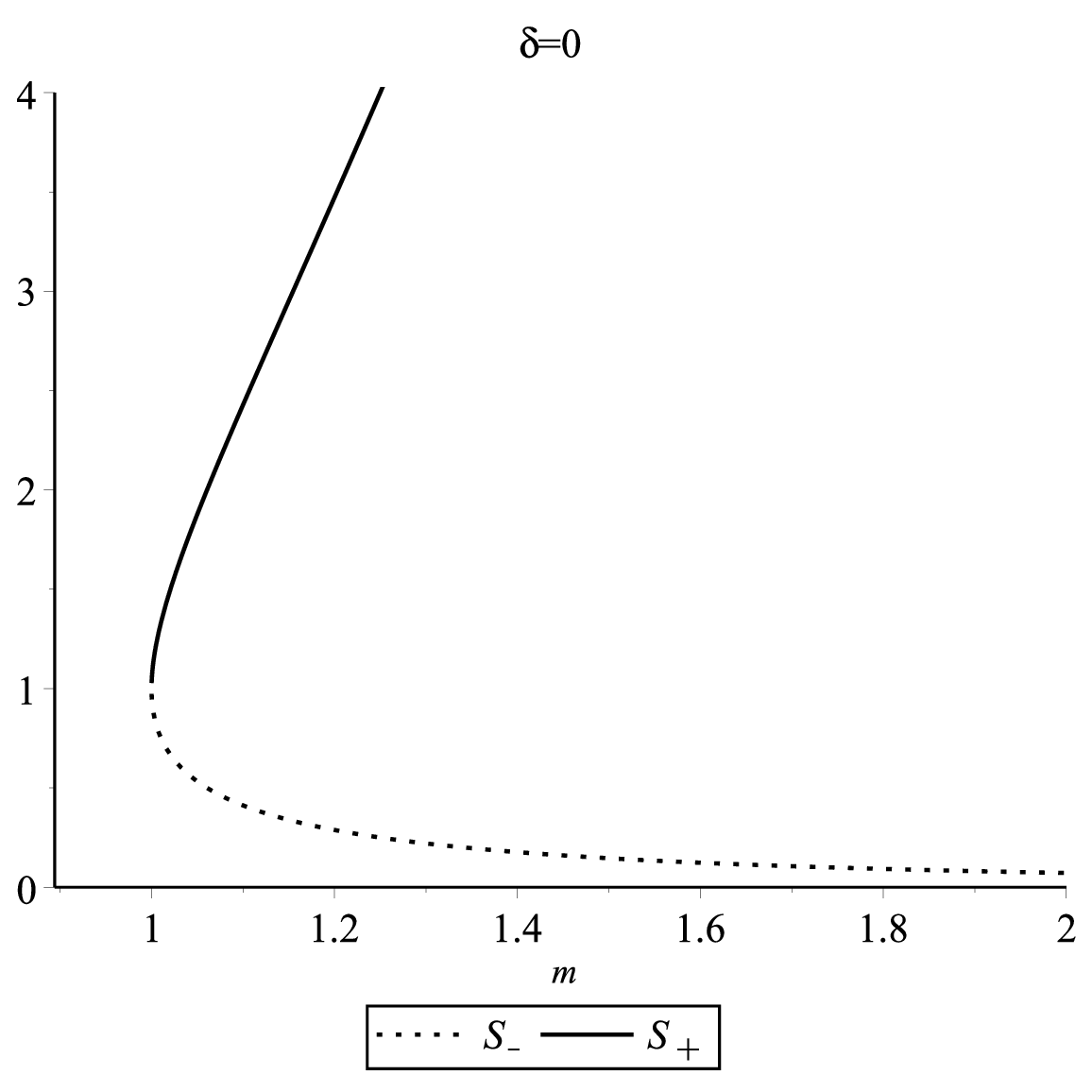}
    \includegraphics[width=6cm,height=6cm]{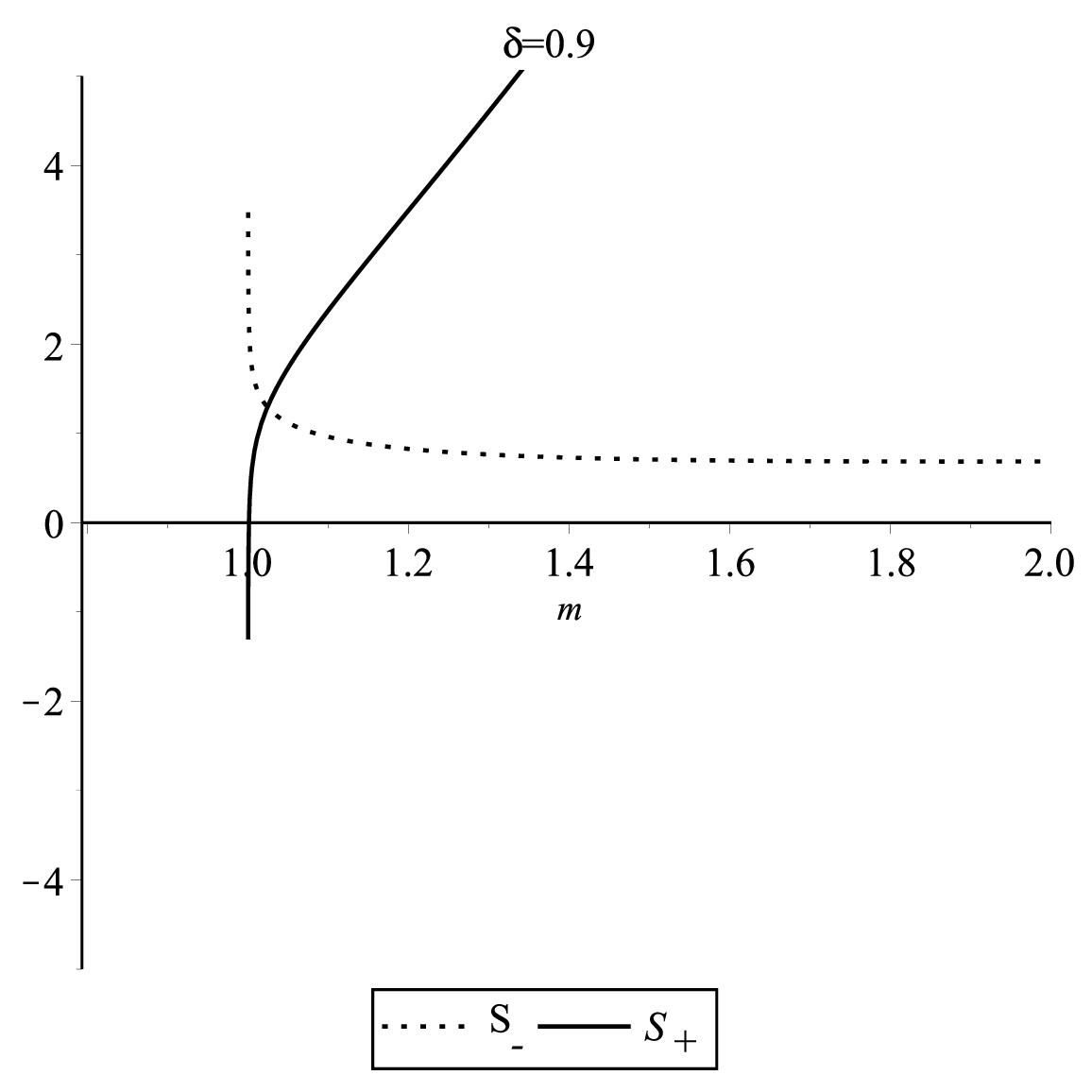}
\includegraphics[width=6cm,height=6cm]{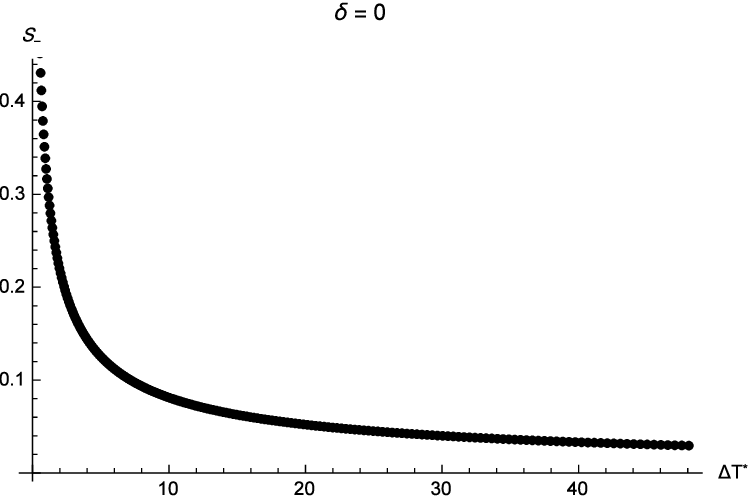}
\includegraphics[width=6cm,height=6cm]{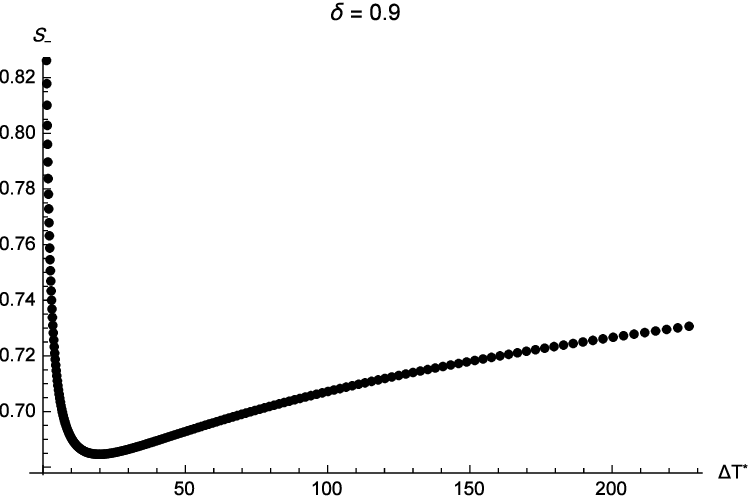}
\includegraphics[width=6cm,height=6cm]{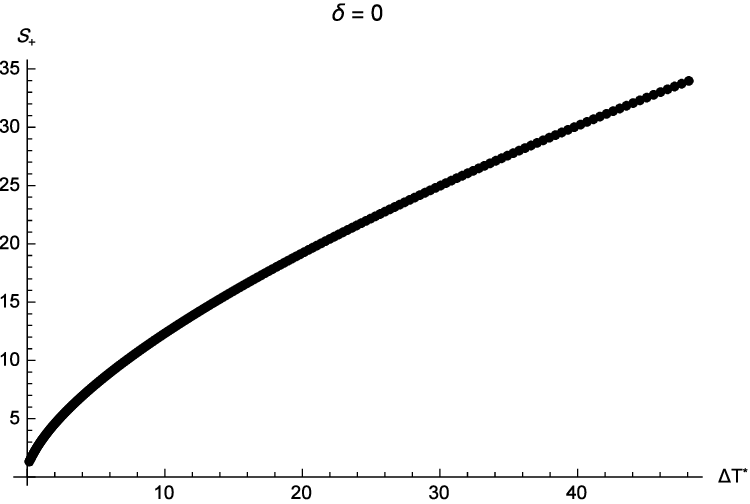}
\includegraphics[width=6cm,height=6cm]{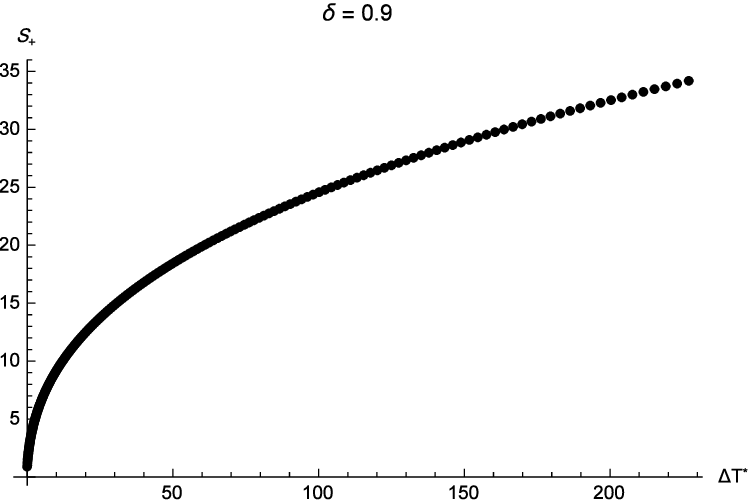}
\caption{\label{fig:epsart} Diagram of $S_{\pm}$ are plotted
against $m$ and $\Delta T^*$ for $\delta=0$ and $\delta=0.9$. }
\end{figure}
\begin{figure}[tbp] \centering
 \includegraphics[width=6cm,height=6cm]{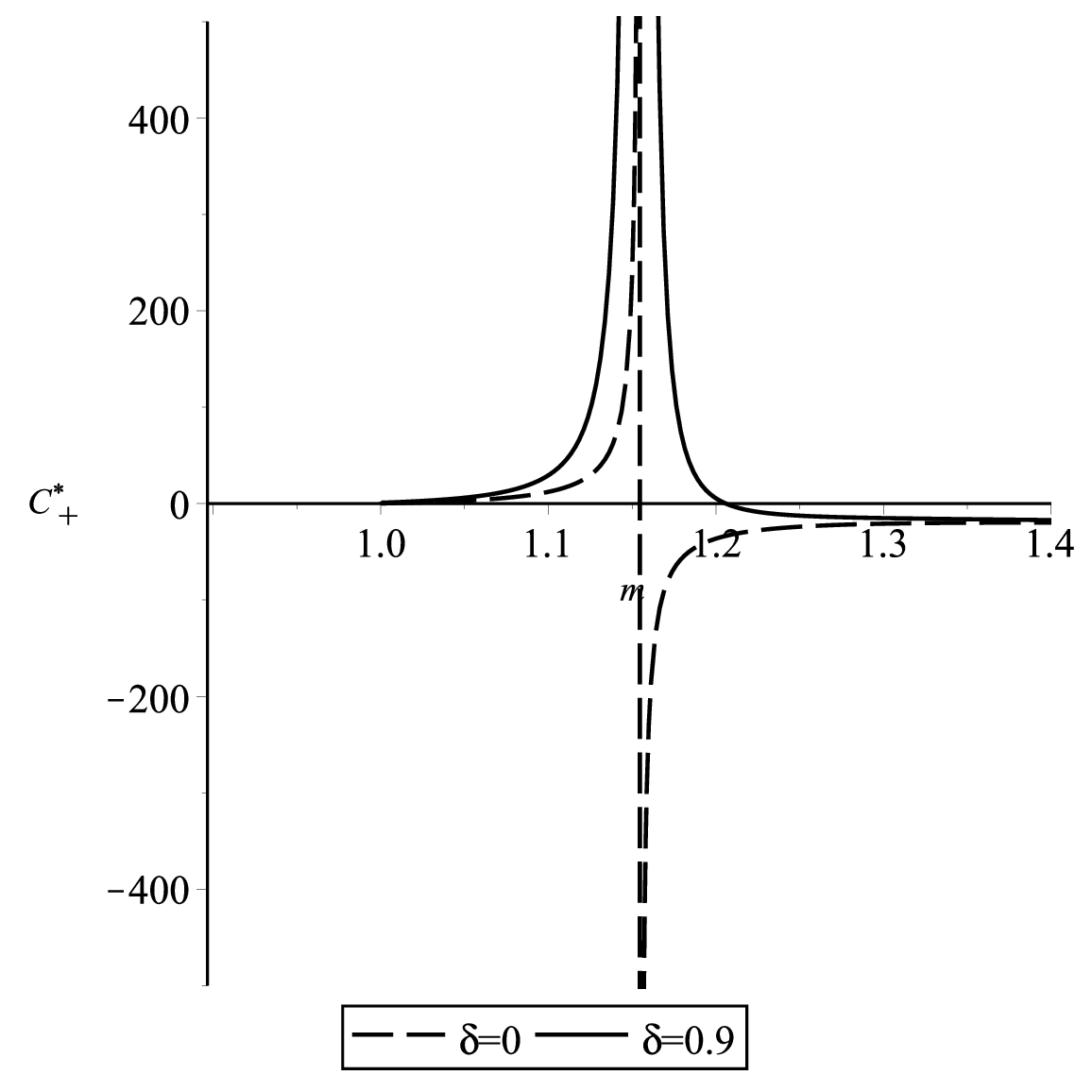}
    \includegraphics[width=6cm,height=6cm]{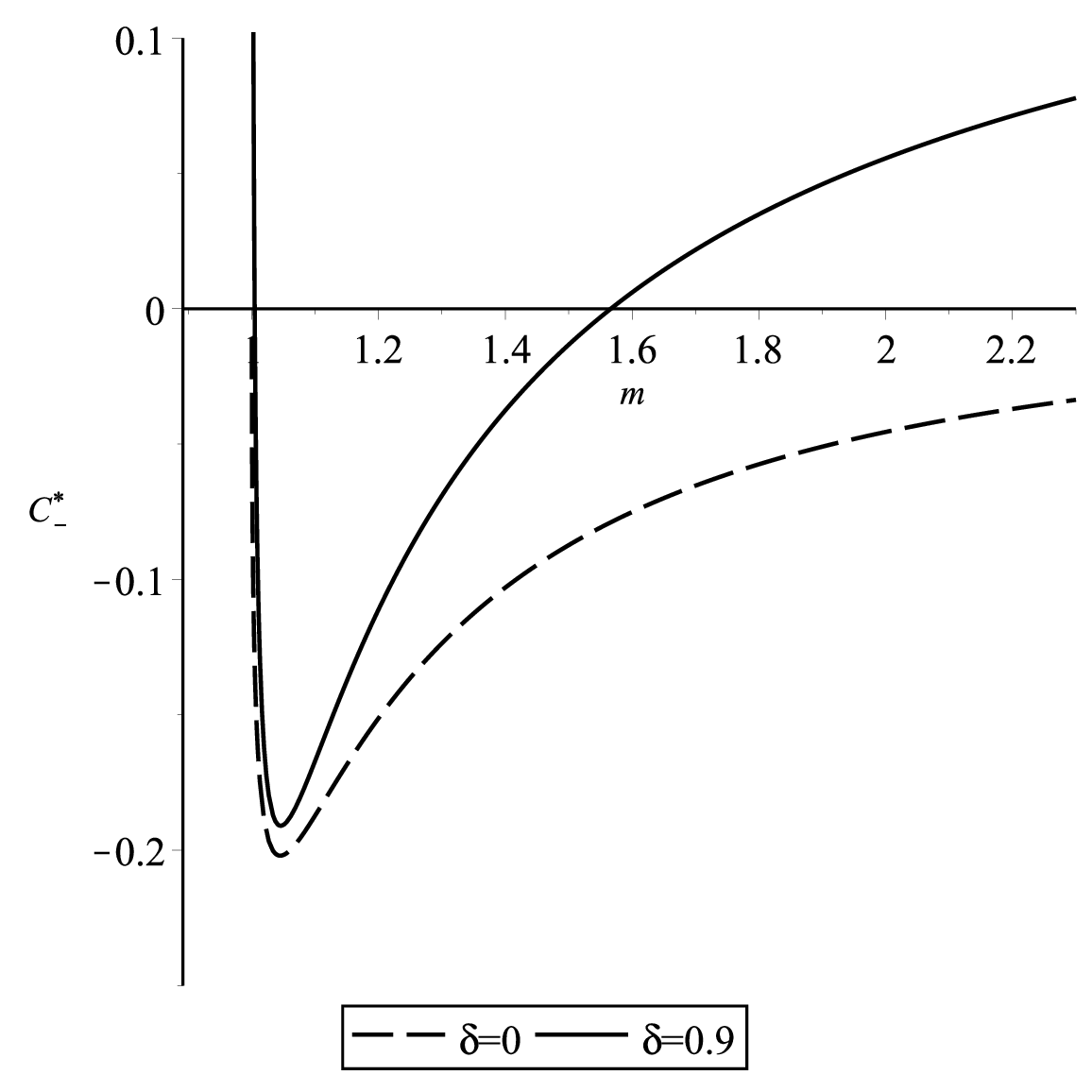}
 \includegraphics[width=6cm,height=6cm]{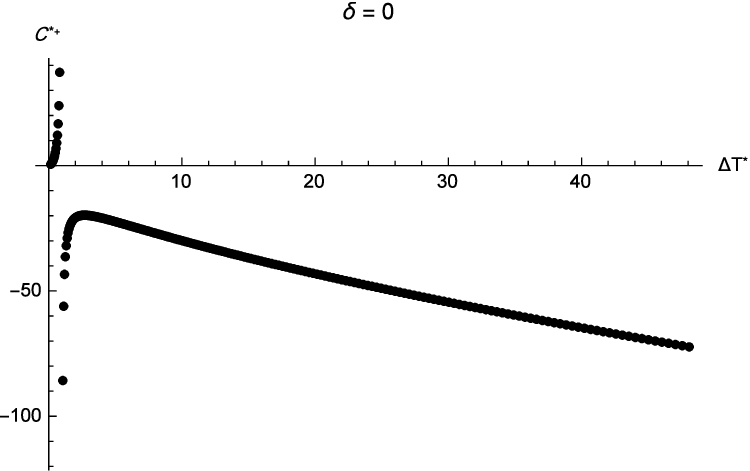}
\includegraphics[width=6cm,height=6cm]{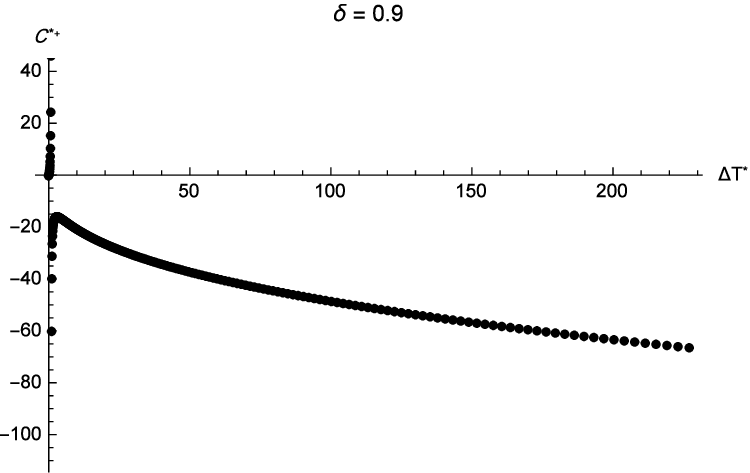}
 \includegraphics[width=6cm,height=6cm]{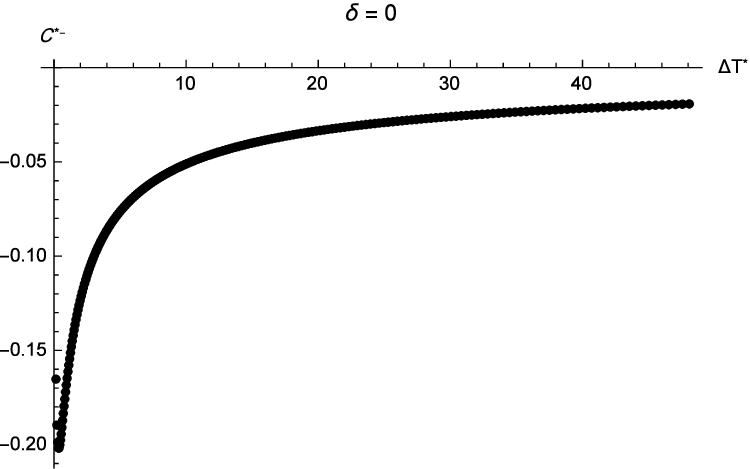}
    \includegraphics[width=6cm,height=6cm]{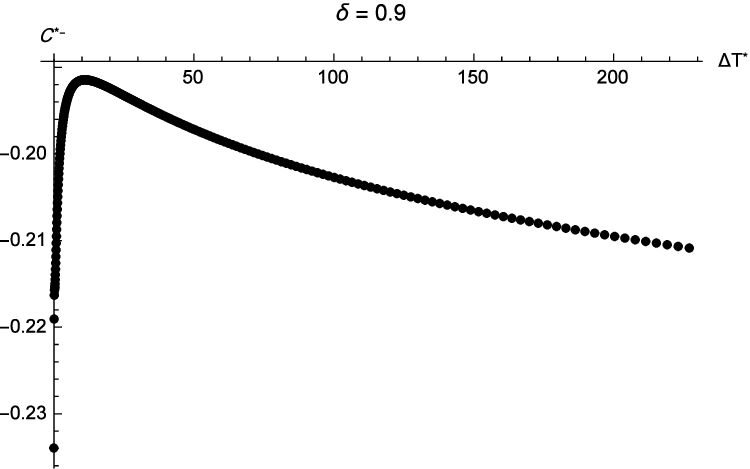}
\caption{\label{fig:epsart} Diagram of interior and exterior
horizons heat capacities $C^*_{\pm}$ are plotted against $m$  and
$\Delta T^*$ for $\delta=0$ and $\delta=0.9$. }
\end{figure}
\begin{figure}[tbp] \centering
\includegraphics[width=6cm,height=6cm]{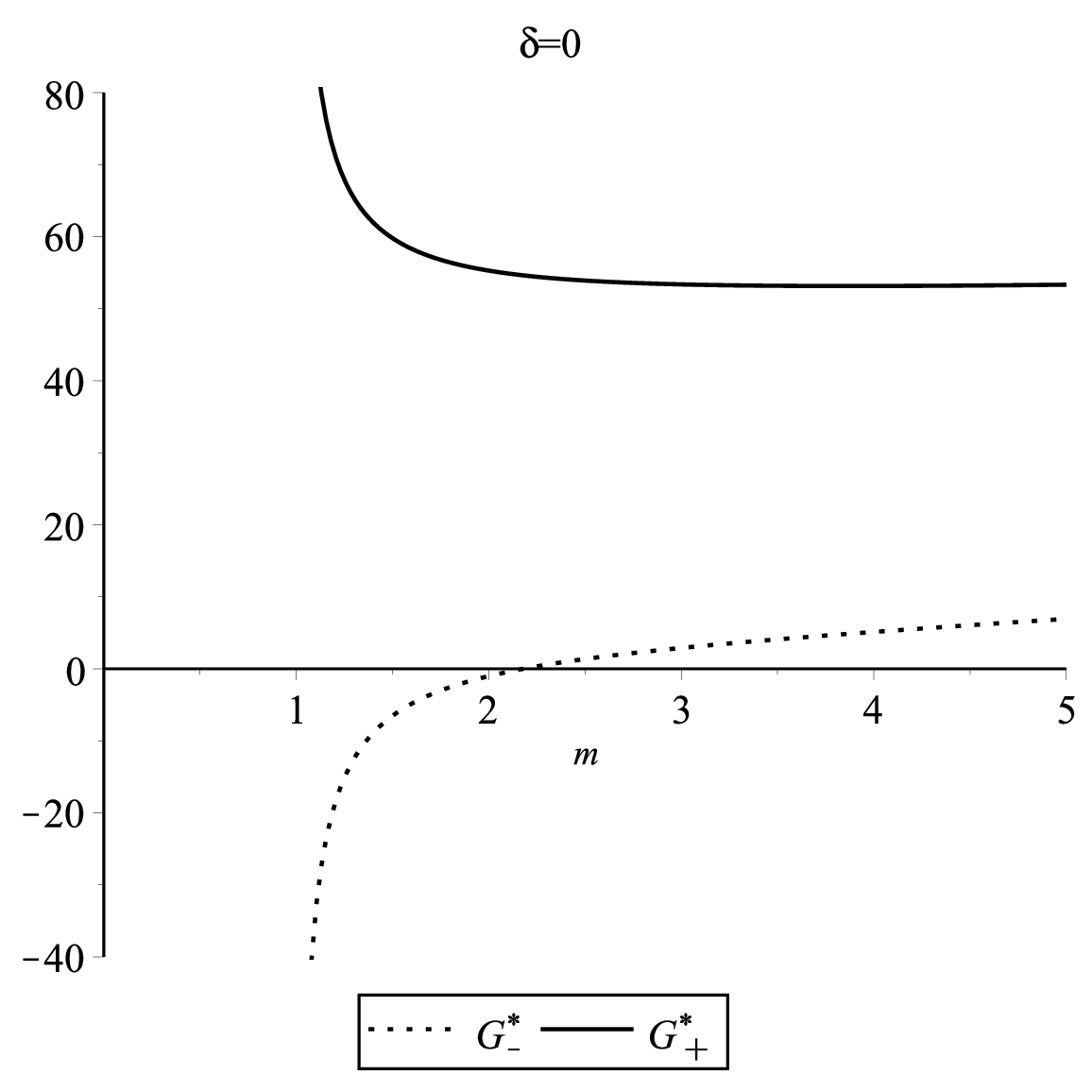}
    \includegraphics[width=6cm,height=6cm]{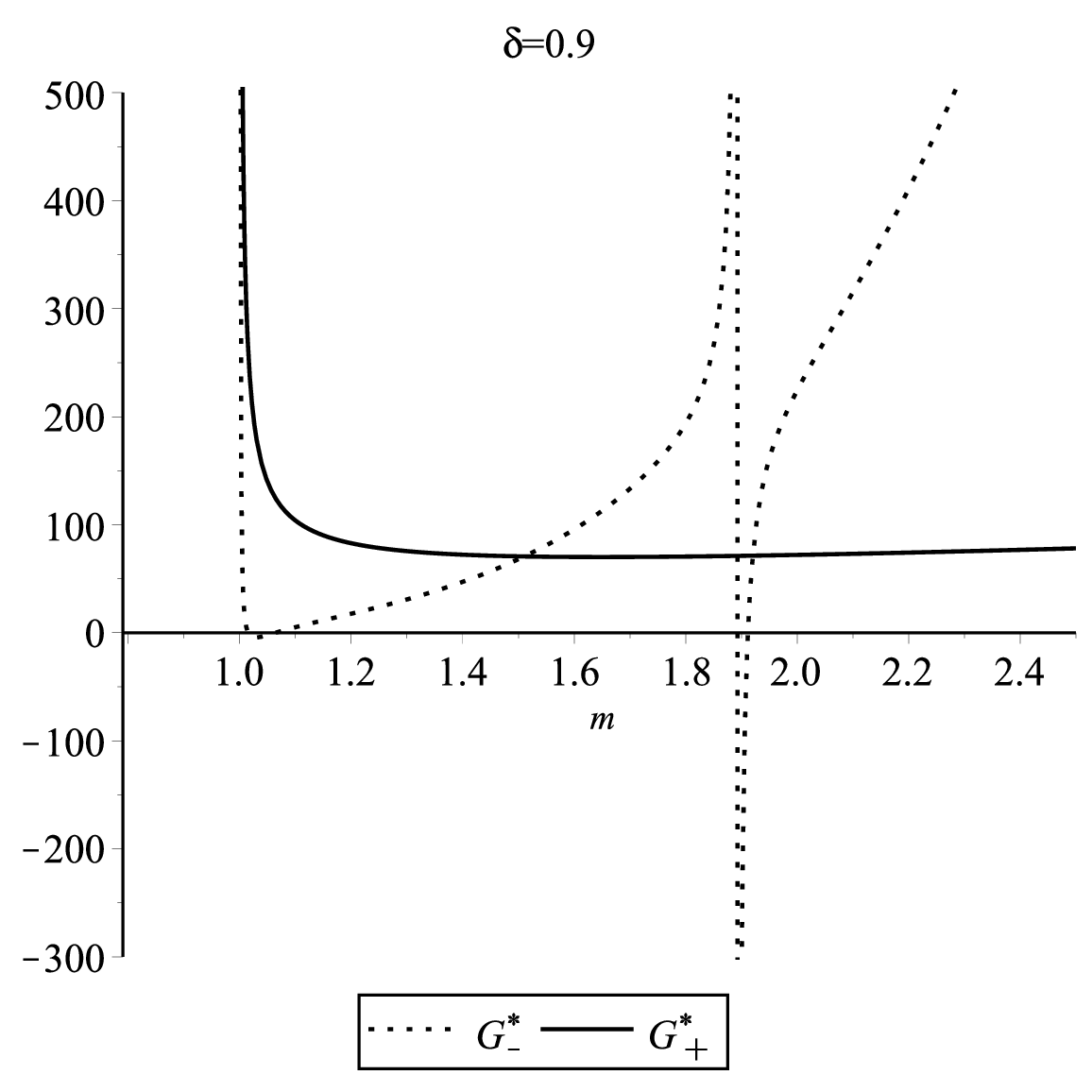}
\includegraphics[width=6cm,height=6cm]{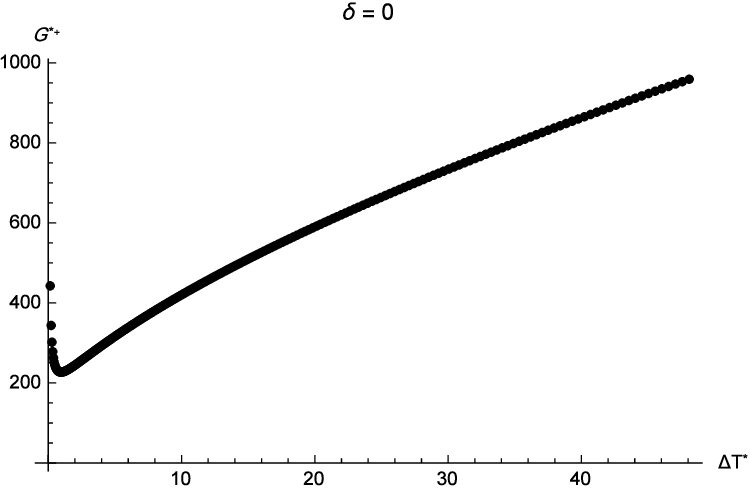}
\includegraphics[width=6cm,height=6cm]{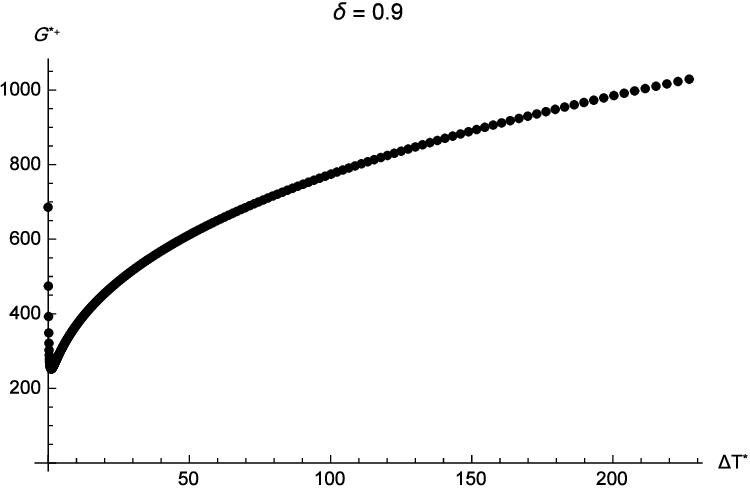}
\includegraphics[width=6cm,height=6cm]{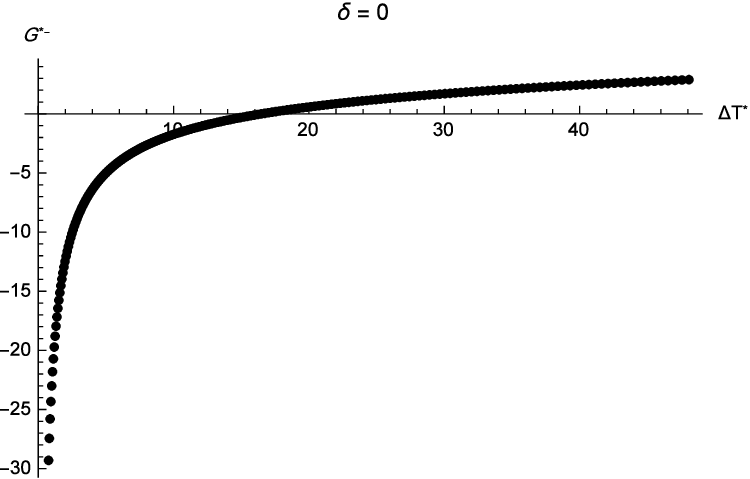}
\includegraphics[width=6cm,height=6cm]{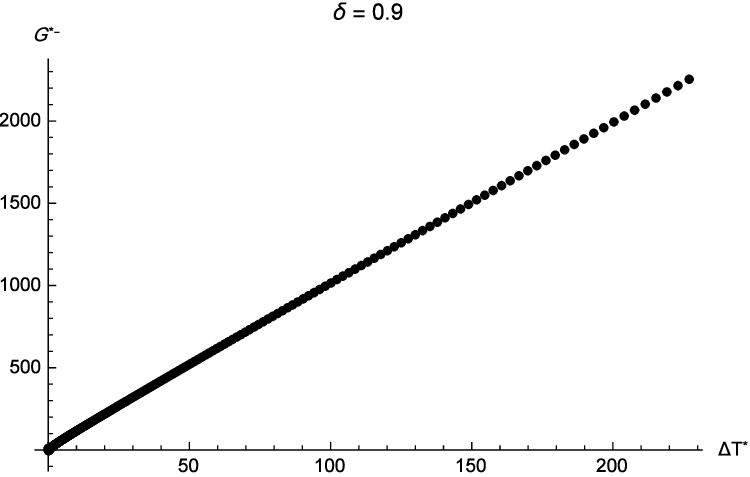}
\caption{\label{fig:epsart} Diagram of interior and exterior
horizons Gibbs free energies $G^*_{\pm}$ are plotted against $m$
and $\Delta T^*$ for $\delta=0$ and $\delta=0.9.$}
\end{figure}
\begin{figure}[tbp] \centering
\includegraphics[width=6cm,height=6cm]{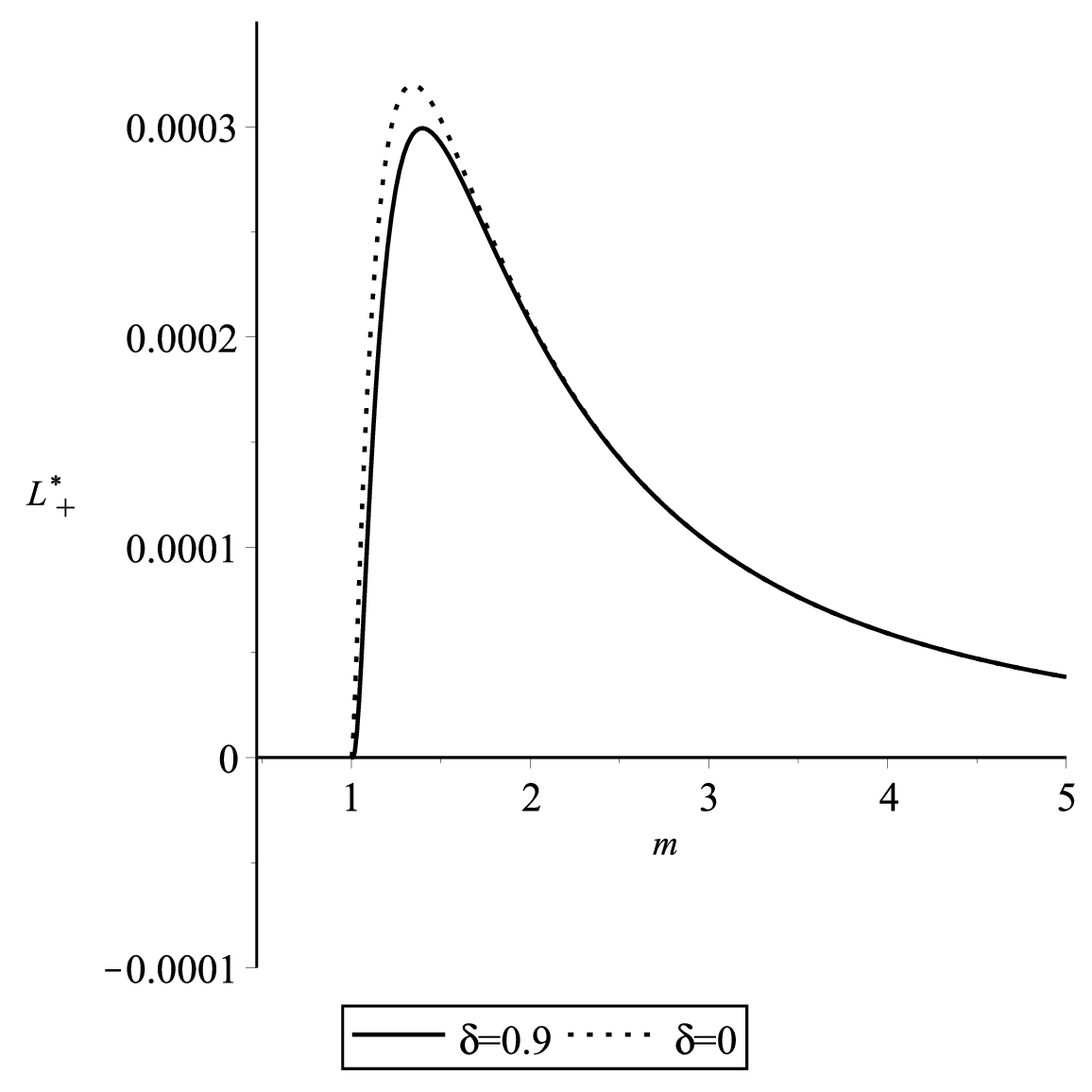}
   \includegraphics[width=6cm,height=6cm]{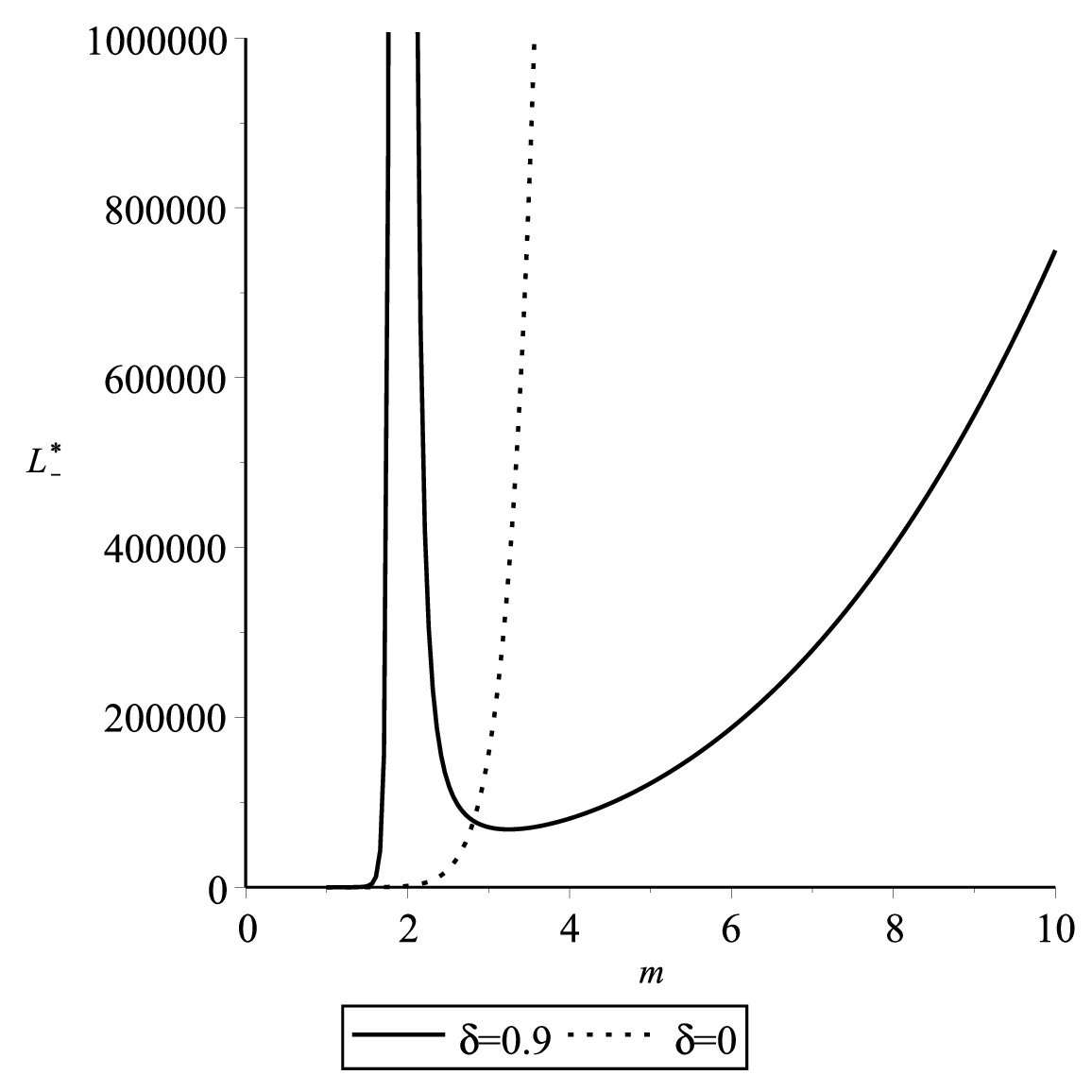}
\includegraphics[width=6cm,height=6cm]{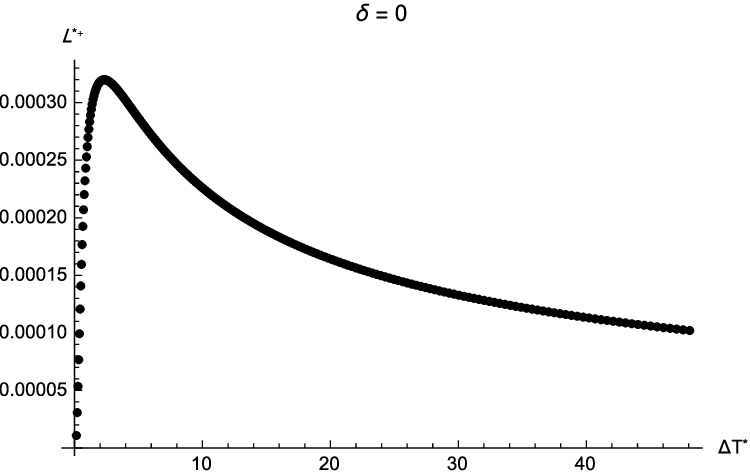}
\includegraphics[width=6cm,height=6cm]{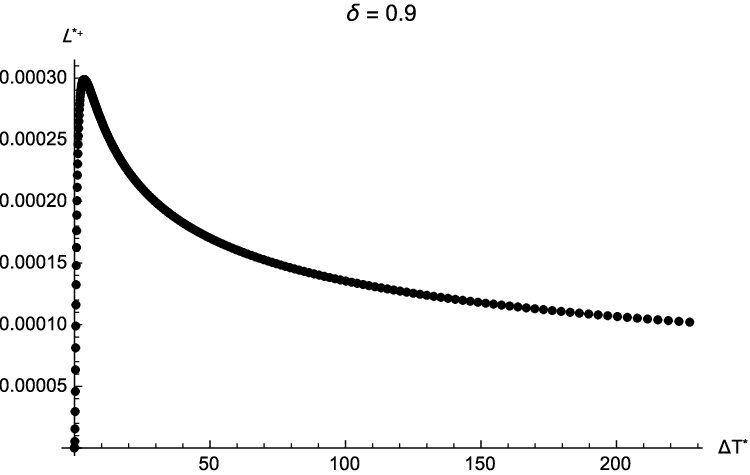}
   \includegraphics[width=6cm,height=6cm]{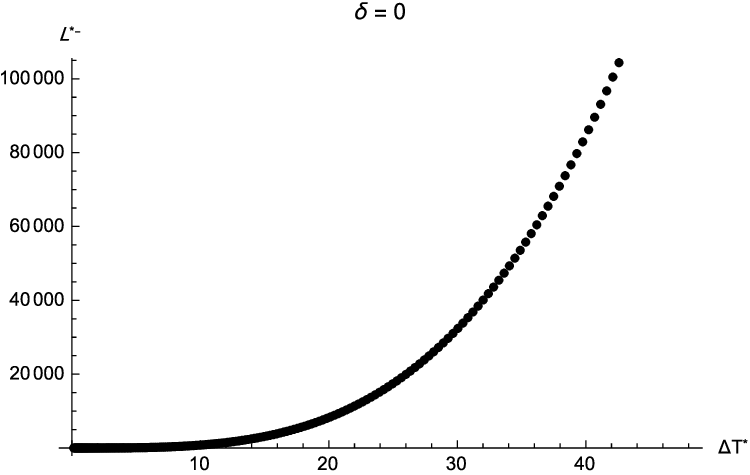}
 \includegraphics[width=6cm,height=6cm]{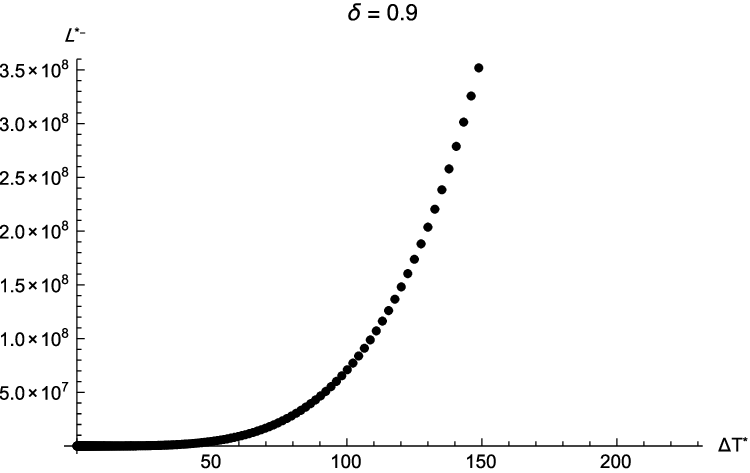}
\caption{\label{fig:epsart} Diagram of exterior and interior
horizons luminosity $L^*_{\pm}$  are plotted against $m$ and
$\Delta T^*$ for $\delta=0$ and $0.9.$}
\end{figure}
\begin{figure}[tbp] \centering
   \includegraphics[width=6cm,height=6cm]{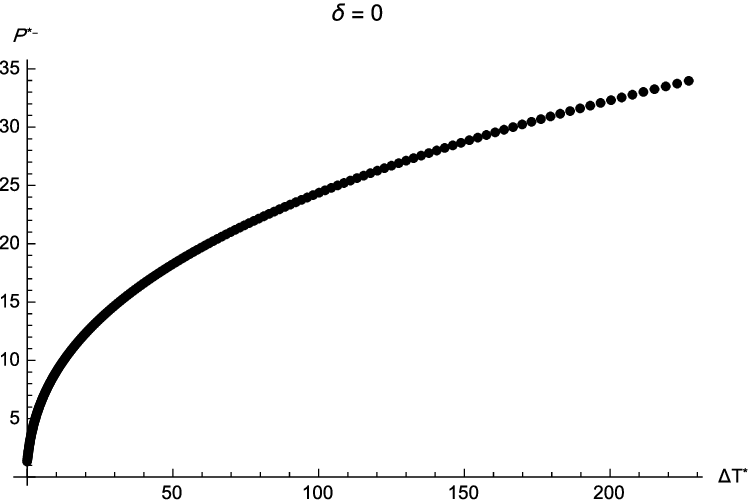}
\includegraphics[width=6cm,height=6cm]{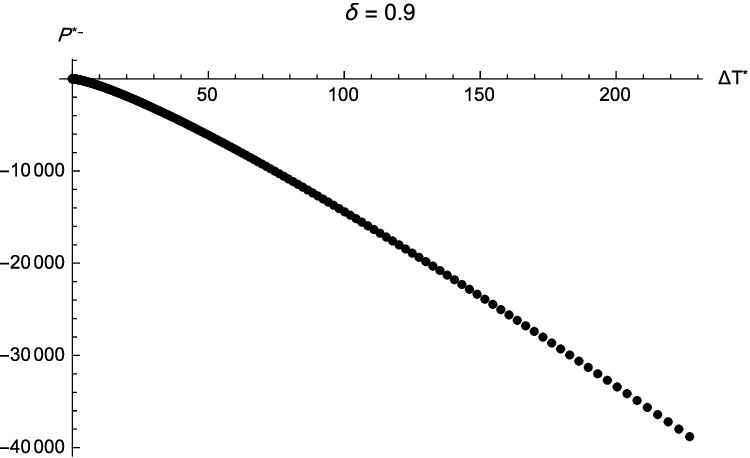}
\includegraphics[width=6cm,height=6cm]{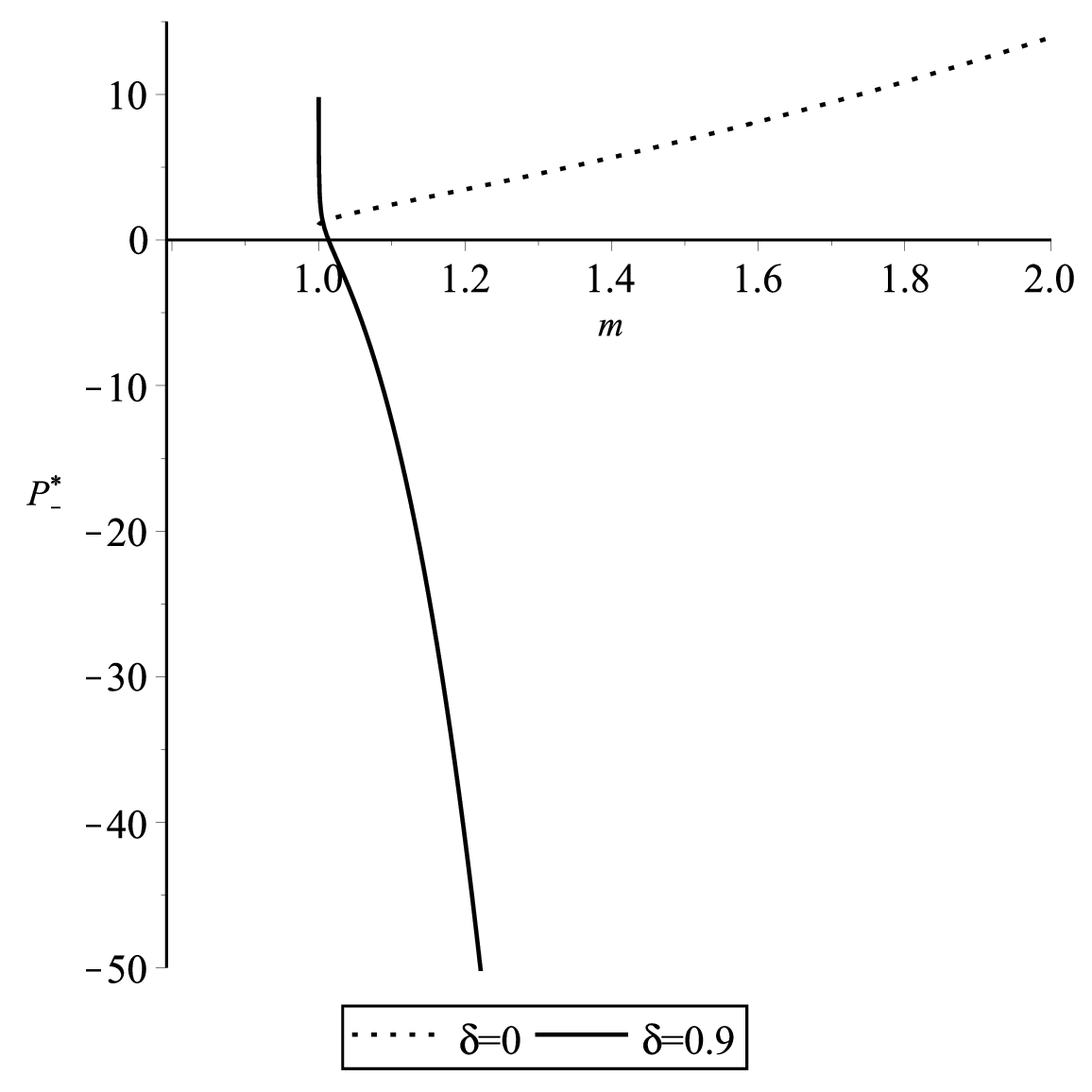}
\caption{\label{fig:epsart} Diagram of interior horizon pressure
$P^*_-$ is plotted against $m$ and $\Delta T^*$ for $\delta=0$ and
$\delta=0.9$. }
\end{figure}
\end{document}